%% file: mdmonojets_paper.tex
\pdfoutput=1
\documentclass[a4paper,11pt]{article}
\usepackage{jheppub,slashed}
\usepackage[utf8]{inputenc}

\newcommand{\beq}{\begin{eqnarray}}
\newcommand{\eeq}{\end{eqnarray}}
\newcommand{\be}{\begin{equation}}
\newcommand{\ee}{\end{equation}}

\def\ltap{\ \raise.3ex\hbox{$<$\kern-.75em\lower1ex\hbox{$\sim$}}\ }
\def\gtap{\ \raise.3ex\hbox{$>$\kern-.75em\lower1ex\hbox{$\sim$}}\ }

\def\eg{{\it e.g.}}
\def\ie{{\it i.e.}}

\def\bsp#1\esp{\begin{split}#1\end{split}} 
\def\bea{\begin{eqnarray}}
\def\eea{\end{eqnarray}}

\definecolor{red1}{cmyk}{0,1,1,0.3}

\newcommand{\amc}{{\sc MadGraph5\textunderscore}a{\sc MC@NLO}}

\title{Monojet searches for momentum-dependent dark matter interactions}

\author[a]{Daniele~Barducci}
\author[b]{, Aoife~Bharucha}
\author[c]{, Nishita~Desai}
\author[d]{, Michele~Frigerio}
\author[e,f]{, Benjamin~Fuks}
\author[e,f,g]{, Andreas~Goudelis}
\author[g]{, Suchita~Kulkarni}
\author[h]{, Giacomo~Polesello}
\author[i]{and Dipan~Sengupta}

\affiliation[a]{LAPTh, Universit\'e Savoie Mont Blanc, CNRS, B.P. 110, F-74941 Annecy-le-Vieux, France
}
\affiliation[b]{CNRS, Aix Marseille U., U. de Toulon, CPT, Marseille, France
}
\affiliation[c]{Institut f\"ur Theoretische Physik, Universit\"at Heidelberg, Germany
}
\affiliation[d]{Laboratoire Charles Coulomb (L2C), UMR 5221 CNRS-Universit\'e de Montpellier, F-34095 Montpellier, France
}
\affiliation[e]{Sorbonne Universit\'es, UPMC Univ.~Paris 06, UMR 7589, LPTHE,
   F-75005 Paris, France}
\affiliation[f]{CNRS, UMR 7589, LPTHE, F-75005 Paris, France}
\affiliation[g]{Institute of High Energy Physics, Austrian Academy of Sciences, Nikolsdorfergasse 18, 1050 Vienna, Austria
}
\affiliation[h]{INFN, Sezione di Pavia, via Bassi 6, 27100 Pavia, Italy
}
\affiliation[i]{Laboratoire de Physique Subatomique et de Cosmologie, Universit\'e Grenoble-Alpes, CNRS/IN2P3, 53 Avenue des Martyrs, F-38026 Grenoble, France
}
\emailAdd{daniele.barducci@lapth.cnrs.fr}
\emailAdd{aoife.bharucha@cpt.univ-mrs.fr}
\emailAdd{n.desai@thphys.uni-heidelberg.de}
\emailAdd{michele.frigerio@umontpellier.fr}
\emailAdd{fuks@lpthe.jussieu.fr}
\emailAdd{andreas.goudelis@lpthe.jussieu.fr}
\emailAdd{Suchita.Kulkarni@oeaw.ac.at}
\emailAdd{giacomo.polesello@cern.ch}
\emailAdd{dipan@lpsc.in2p3.fr}

\abstract{
We consider minimal dark matter scenarios featuring
momentum-dependent couplings of the dark sector to the Standard Model.
We derive constraints from existing LHC searches in the monojet channel, estimate the future
LHC sensitivity for an integrated luminosity of 300 fb$^{-1}$, and compare 
with models exhibiting conventional momentum-independent
interactions with the dark sector.
In addition to being well motivated by (composite) pseudo-Goldstone dark matter scenarios, momentum-dependent couplings are interesting as they weaken direct detection constraints.
 For a specific dark matter mass, the LHC turns out to
be sensitive to smaller signal cross-sections in the momentum-dependent case, by
virtue of the harder jet transverse-momentum distribution.
}

\begin{document}
\begin{flushright}

    \hspace{3cm}LAPTH-041/16 \\

\end{flushright}

\maketitle
\flushbottom

\section{Introduction} \input{1-intro}

\section{Theoretical framework and constraints} \input{2-model}

\section{Numerical results}\input{3-results}
\section{Conclusion}\input{4-conclusions}

\section{Acknowledgements}
The authors are grateful to F.~Maltoni for discussions about the unitarity
constraints and to S.~Lacroix for his fruitful participation to the early
stage of this project. We would also like to thank the organizers of the 2015
`Les Houches -- Physics at TeV colliders' workshop where this work was
initiated.

MF has been partly supported by the OCEVU Labex (ANR-11-LABX-0060) and the
A*MIDEX project (ANR-11-IDEX-0001-02) and by the European Union's Horizon 2020
research and innovation programme under the Marie Sk\"odowska-Curie grant
agreements No 690575 and No 674896.
BF has been supported by the Theorie LHC France initiative of the CNRS, SK by
the New Frontiers program of the Austrian Academy of Sciences, DS by the French
ANR project DMAstroLHC (ANR-12-BS05-0006). 


\appendix
\section{Derivation of perturbative unitarity constraints}
\input{5-unitarity}

\bibliographystyle{JHEP}
\bibliography{mdmonojets_paper}

\end{document}

%% file: 1-intro.tex
Collider searches for final states consisting of a hard jet and missing
energy~\cite{Aad:2014nra,Aad:2015zva,Aaboud:2016tnv,Khachatryan:2014rra,Khachatryan:2015wza},
dubbed monojet searches, provide a means to detect new invisible particles that
are stable on detector or even cosmological scales. In the latter case, these
particles could contribute to the dark matter (DM)
energy density of the Universe and monojet searches could offer
invaluable information about their existence.
Furthermore it is well known that the jet transverse-momentum spectrum is one of the
key observables that could unravel the nature of the dark
matter couplings to the Standard Model from monojet
probes~\cite{Rizzo:2008fp, Buchmueller:2014yoa}.
In this work, we study the effect of
derivative and non-derivative couplings between the Standard Model
and the new physics sector on the monojet kinematics.
Our preliminary results, including only 8 TeV LHC data, appeared in the proceedings of the ``Les Houches 2015 -- Physics at TeV colliders'' workshop~\cite{Brooijmans:2016vro}.

Models with derivative couplings are motivated by new physics setups
featuring pseudo-Nambu-Goldstone bosons (pNGBs), \ie~light scalar fields
connected to the spontaneous breaking of a global symmetry at an energy scale
$f$. More concretely, this class of models includes composite Higgs scenarios
where the set of pNGBs involves the Higgs boson and possibly extra dark scalar
particles~\cite{Frigerio:2012uc,Marzocca:2014msa, Fonseca:2015gva,Brivio:2015kia,Bruggisser:2016nzw,Bruggisser:2016ixa}. In this case, the pNGB shift symmetry indeed only allows
for derivative (momentum-dependent) pNGB interactions suppressed by powers of
the scale $f$. An explicit weak breaking of the shift symmetry, parameterized by
a small coupling strength $\epsilon$, is however necessary in order to induce
pNGB masses, which subsequently generates additional non-derivative
momentum-independent couplings
proportional to $\epsilon/f$. In this work, we rely on a simplified effective field theory
approach where the form of the Lagrangian is inspired by such pNGB setups, with all
specific and model-dependent assumptions for the new physics masses and
couplings being, however, relaxed. As already pointed out in the literature, this
effective Lagrangian approach is appropriate for interpreting LHC missing energy
signatures within frameworks featuring light dark matter particles interacting
with the Standard Model via non-renormalizable derivative
operators~\cite{Bruggisser:2016ixa,Bruggisser:2016nzw}.

Most ultraviolet-complete models of dark matter involve additional particles,
potentially carrying Standard Model quantum numbers. Although dedicated LHC
searches could detect such additional states, we consider a simple setup where
the only new states that are accessible at the LHC are the dark matter particle
and (sometimes) the mediator connecting the Standard Model and the dark sector.
More specifically, we first consider an invisible sector solely comprised of a
Standard Model-singlet real scalar field $\eta$ which is taken to be odd
under a ${\cal Z}_2$ discrete symmetry. The Standard Model fields are then
chosen to be even under the same ${\cal Z}_2$ symmetry, which forbids the decay of the
$\eta$ particle into any ensemble of Standard Model states and renders it a
potential dark matter candidate. In a minimal scenario, the mediator is taken
to be the Standard Model Higgs field $H$ that interacts with $\eta$ via both a
renormalizable quartic coupling and a non-renormalizable derivative coupling.
This framework, however, turns out to be strongly constrained by LHC measurements of
the Higgs boson properties. As an alternative we therefore consider a slightly extended
setup where we additionally introduce a real gauge-singlet ${\cal Z}_2$-even 
scalar mediator particle $s$, a choice which allows us to avoid these
constraints. While $s$ couples to the dark sector both through a derivative (dimension-five) 
and a non-derivative renormalizable operator, it is connected to the Standard Model only 
through a (potentially loop-induced) dimension-five operator involving gluon field strength bilinears.
This simple model not only reproduces the observed dark matter abundance of the universe but also, assuming that momentum-dependent interactions dominate, can evade the direct detection constraints.

In this paper, we first provide details of our theoretical framework in
Section~\ref{sec:theo} and then examine the LHC constraints stemming from the
monojet analysis performed by the ATLAS collaboration for proton-proton collisions at a center-of-mass energy of 13 TeV with 3.2 fb$^{-1}$ in Section~\ref{sec:mdmonojets_lhc}. We assess the effects of
momentum-dependent and momentum-independent dark matter couplings on monojet
distributions and derive the corresponding bounds for both cases. Dijet searches for the mediator $s$ at
past and present colliders are also taken into account and discussed, and we
finally entertain the possibility that the $\eta$ particle is responsible for
the measured dark matter density in the Universe. In this spirit, we
investigate the dependence of the relic abundance and the direct dark matter
detection constraints on the model parameters. Our findings are summarized in
Section~\ref{sec:conclusion}, and technical details on the
range of validity of our effective description based on perturbative unitarity arguments are presented in
Appendix~\ref{pertunitarity}.

%% file: 2-model.tex
\label{sec:theo}

\subsection{The minimal scenario: the Higgs portal}

In order to study the impact of derivative and non-derivative couplings of dark
matter to the Standard Model, we first consider a minimal setup
involving both momentum-dependent and momentum-independent couplings of the dark
matter particle. We impose that the dark matter only couples to the Higgs field, which plays
the role of the mediator.

We supplement the Standard Model by a gauge-singlet real scalar field $\eta$
that is chosen odd under a ${\cal Z}_2$ symmetry, where in contrast the Standard Model fields
are taken to be even. The $\eta$ particle then only interacts with the
Standard Model through couplings to the Higgs doublet $H$, such that the model
Lagrangian reads
\be
{\cal L}_\eta = {\cal L}_{\rm SM} +
\frac 12 \partial_\mu\eta \partial^\mu\eta -\frac 12 \mu_\eta^2 \eta^2 -\frac 14 \lambda_\eta \eta^4 - \frac 12 \lambda \eta^2 H^\dagger H 
+\frac {1}{2f^2} (\partial_\mu \eta^2) \partial^\mu(H^\dagger H)~.
\label{eq:mdmonojets_lag}
\ee
This Lagrangian contains renormalizable operators compatible with the ${\cal Z}_2$ symmetry
(\mbox{$\eta\rightarrow -\eta$}) and a di\-men\-si\-on-six operator involving
derivatives. While several other non-derivative dimension-six operators are
allowed by the model symmetries, these are not expected to have a significant
impact on the monojet analysis. As the effect of these operators is
negligible for our purposes, we have omitted these in our
parameterization of Eq.~\eqref{eq:mdmonojets_lag}. In the context of composite
Higgs models, the scalar field $\eta$ may be a pNGB and $f$
its decay constant. The theoretical motivations for this minimal model and the resulting dark matter phenomenology
are described in Ref.~\cite{Frigerio:2012uc}. Further related studies
are also available in the literature~\cite{Marzocca:2014msa, Fonseca:2015gva,Brivio:2015kia}.

After electroweak symmetry breaking, the part of the Lagrangian containing the interactions of $\eta$ with the physical Higgs boson $h$ is given by
\be
{\cal L}_\eta \supset 
- \frac 14 (v+h)^2 \left( \lambda \eta^2 + \frac {1}{f^2} \partial_\mu\partial^\mu\eta^2 \right)~,
\label{eq:mdmonojets_int1}\ee
and the $\eta$ mass $m_\eta$ satisfies
\be
 m_\eta^2 = \mu_\eta^2 + \lambda v^2/2~.
\ee
The trilinear scalar interaction in Eq.~\eqref{eq:mdmonojets_int1} yields monojet events at the LHC via, for instance, the gluon fusion process \mbox{$gg\to g h^{(*)}\to g\eta\eta$}, while the quartic
interactions give rise to mono-Higgs events \mbox{$gg\to h^* \to  h \eta\eta$}.
When $2m_\eta < m_h$, the Higgs boson is produced on-shell and the strength of
the derivative interaction vertex is proportional to $p_h^2/f^2 =m_h^2/f^2$,
with $p_h$ being the final-state Higgs boson four-momentum. The
momentum-dependence reduces to a constant, so that momentum-dependent interactions
become indistinguishable from their momentum-independent counterparts. In this
regime, bounds
from monojet searches are found to be weaker than the constraints stemming from the
Higgs invisible width results~\cite{Aad:2015pla,Khachatryan:2015vta, Khachatryan:2014jba},
\be
\Gamma(h\rightarrow \eta\eta)=\frac{v^2}{32\pi m_h}\left(\frac{m_h^2}{f^2} - \lambda\right)^2 \sqrt{1-\dfrac{4m_\eta^2}{m_h^2}} ~\theta(m_h^2-4m_\eta^2) 
\lesssim 0.15 \Gamma_h^{\rm SM} \simeq  0.7{\rm~MeV} ~,
\label{eq:mdmonojets_invC}
\ee
at the 95\% confidence level.

We therefore focus on the complementary kinematic region
where
\mbox{$2m_\eta > m_h$}. The monojet
signal arises from off-shell Higgs-boson production, and the derivative
interactions of the $\eta$ particle result in a strong momentum dependence at
the differential cross-section level. The subsequent differences in the jet
transverse momentum distribution could allow us to discriminate derivative from non-derivative dark
matter couplings. This however comes at the
price of a suppression
of the monojet signal, since the relevant partonic cross-section $\hat \sigma$
depends on the virtuality of the Higgs boson $p_h^2$ as
\be
\hat\sigma(gg\rightarrow gh^*\rightarrow g\eta\eta) \propto
\frac{\theta(p_h^2-4m_\eta^2)}{(p_h^2-m_h^2)^2+\Gamma_h^2 m_h^2} \left(\dfrac{p_h^2}{f^2} - \lambda\right)^2\sqrt{1-\dfrac{4m_\eta^2}{p_h^2}} ~,
\ee
where $\Gamma_h$ denotes the Higgs total width. The denominator is clearly
larger in the region where the Higgs is off-shell, or equivalently when \mbox{$p_h^2>4m_\eta^2>m_h^2$}.
\\\vspace{.5cm}

A preliminary monojet analysis within the considered theoretical framework has
recently been performed~\cite{lacroix}, and the collider signatures of this
off-shell Higgs portal model are discussed in Ref.~\cite{Craig:2014lda}.
Our numerical analysis however indicates that in the
light of current experimental data, the
monojet signal is too weak to be observed at the LHC. The precise determination
of the Higgs-boson mass and the important LHC constraints on its production
cross-section and decay width indeed result in tight bounds on the free
parameters of the model,
\be
  m_\eta \gtrsim m_h/2,\qquad \lambda\lesssim 1, \qquad
  f\gtrsim 500~{\rm GeV} - 1~{\rm TeV}\ ,
\ee
where the latter bound applies to models in which the Higgs is a composite pNGB.
As a consequence, the total monojet cross-section after including a selection on the jet
transverse momentum of $p_T^{\rm jet} > 20$~GeV is always smaller than 1~fb and
0.5~fb when only momentum-dependent and momentum-independent couplings are
allowed, respectively, for a center-of-mass energy of \mbox{$\sqrt{s} = 13$~TeV}.

\subsection{A pragmatic scenario with a scalar singlet mediator}\label{sec:mdmonojets_ourmodel}

Given the tight constraints discussed in the previous section, we extend our
framework to analyse a scenario less severely constrained by data. In addition to the dark matter field $\eta$ we
introduce a second scalar mediator field $s$, chosen to be even under the
${\cal{Z}}_2$ symmetry and a singlet under the Standard Model gauge symmetries. We
moreover impose that the scalar potential does not spontaneously break the
${\cal{Z}}_2$ symmetry, or equivalently that $\eta$ does not acquire a
non-vanishing vacuum expectation value (vev). We further demand, without
any loss of generality, that the vev of the $s$ field vanishes as the latter
could always be absorbed in a redefinition of the couplings.\\\\
The relevant Lagrangian is given by
\be
\bsp
{\cal L}_{\eta,s} = &\ {\cal L}_{\rm SM} +
\frac{1}{2}\partial_\mu\eta \partial^\mu\eta - \frac{1}{2} m_\eta^2 \eta\eta
+ \frac{1}{2} \partial_\mu s \partial^\mu s - \frac{1}{2} m_s^2 ss \\
& \ + \frac{c_{s\eta}f}{2}  s \eta\eta + \frac{c_{\partial s\eta}}{f} (\partial_\mu s) (\partial^\mu \eta)\eta + \frac{\alpha_s}{16\pi} \frac{c_{sg}}{f} sG^a_{\mu\nu}G^{a\mu\nu}~,
\label{eq:mdmonojets_thelagrangian}
\esp
\ee
where we include an effective coupling between the $s$ and gluon fields.
Consequently, the
mediator can be produced at the LHC via gluon fusion and can give rise to a
monojet signal via the mechanism \mbox{$gg\rightarrow gs^* \to g \eta\eta$}.
In ultraviolet-complete models, this $c_{sg}$ coupling can be generated by
additional new physics. For example, in a scenario featuring a
vector-like color-triplet fermion $\psi$ of mass \mbox{$M_\psi \gg m_s$} that
interacts via a Yukawa interaction $y_\psi \bar{\psi}\psi s$, triangle
loop-diagram contributions induce \mbox{$c_{sg} = (4/3)(y_\psi f/M_\psi)$}.
Finally, the non-derivative coupling $c_{s\eta}$ governs the momentum-independent
interaction of the two scalars $s$ and $\eta$ and, for a given value of the scale $f$, the derivative coupling
$c_{\partial s\eta}$ controls the strength of the leading momentum-dependent
interactions.

The Lagrangian given in Eq.~\eqref{eq:mdmonojets_thelagrangian} only includes
interactions that are relevant to our analysis, and the considered
dimension-five operator is the unique independent derivative dimension-five
operator inducing an interaction between $s$ and $\eta$. The model can hence be described in terms of six parameters,
\begin{equation}
  m_{s},\quad
  m_{\eta},\quad
  f, \quad
  c_{s\eta},\quad
  c_{\partial s\eta}\quad\text{and}\quad
  c_{sg} .
\end{equation}
Strictly speaking, only five of these parameters are independent as one can
always choose \mbox{$c_{\partial s\eta}=1$} and determine the strength of the
momentum-dependent interaction by varying $f$. This choice is motivated by
models where $s$, $\eta$ and the Higgs boson are pNGBs associated with the
spontaneous breaking of a global symmetry at a scale $f$ and where
$c_{\partial s\eta}$ is expected to be of order one. In this case, the $f$ parameter is
constrained by precision Higgs and electroweak data that roughly
imposes \mbox{$f\gtrsim 500~{\rm GeV} - 1~{\rm TeV}$}~\cite{Panico:2015jxa}. In our numerical
analysis of Section~\ref{sec:mdmonojets_lhc}, we consider four
representative values for the $s$ particle mass, $m_s = 50$~GeV, 250~GeV,
500~GeV and 750~GeV, which allows us to cover a wide range of mediator masses.

\subsection{Constraints on the parameters of the model}
The model can be
constrained in a number of ways. In particular, searches for dijet resonances could
be promising since a singly-produced mediator via gluon fusion often decays back
into a pair of jets (\mbox{$g g\to s^{(*)} \to gg$}). For the case where $\eta$
is a viable dark matter candidate, the model should in additional yield a relic
density in agreement
with Planck measurements and satisfy direct dark matter detection bounds. Before
getting into a detailed investigation of these constraints, we first study
some properties of the model in order to understand the bounds that
can be expected from collider, cosmological and theoretical considerations. A
complete set of numerical results is then presented in
Section~\ref{sec:mdmonojets_lhc}.

From the Lagrangian given in Eq.~\eqref{eq:mdmonojets_thelagrangian}, the
partial decay widths of the $s$ particle into gluon and $\eta$ pairs are calculated to be
\begin{align}
\Gamma(s \rightarrow gg) & = \frac{\alpha_s^2 c_{sg}^2 m_s^3}{128 \pi^3 f^2}~, \\
\Gamma(s \rightarrow \eta \eta) 
& = \frac{f^2}{32\pi m_s} \left( c_{\partial s\eta} \frac{m_s^2}{f^2} + c_{s\eta} \right)^2  \sqrt{1 - \frac{4 m_\eta^2}{m_s^2}}\ \theta(m_s^2-4m_\eta^2)~,
\end{align}
in agreement with results obtained using the decay module of
{\sc FeynRules}~\cite{Alwall:2014bza,Alloul:2013bka}. For the coupling
values adopted in our analysis, the total width $\Gamma_s$ is always small. This
implies that the narrow width approximation can be used for any cross-section
calculation involving a resonant $s$-contribution. The $s$-induced dijet
cross section can hence be expressed as
\be
\sigma(pp \rightarrow s \rightarrow gg) = 
  \int_0^1 {\rm d}x_1 \int_0^1{\rm d} x_2 \ f_g(x_1,m_s) f_g(x_2,m_s) 
     \frac{ \alpha_s^2 c_{sg}^2 m_s^2}{1024 \pi f^2} \ \delta(\hat s-m_s^2)
    {\rm BR}\Big(s\to gg\Big)\ ,
\ee
where $\sqrt{\hat{s}}$ denotes the partonic center-of-mass energy and $f_g(x,\mu)$ the
universal gluon density that depends on the longitudinal momentum fraction $x$
of the gluon in the proton and the factorization scale $\mu$. For the considered
values of $m_s$, the most stringent dijet constraints arise from
Sp$\bar{\rm p}$S~\cite{Alitti:1993pn} and Tevatron~\cite{Aaltonen:2008dn} data which
provides upper limits on the new physics cross section $\sigma$ for mediator masses of \mbox{140 -- 300~GeV} and \mbox{200 -- 1400~GeV} respectively.
In comparison, the LHC Run I results further extend the range of covered
mediator masses up to 4.5~TeV~\cite{Khachatryan:2015sja,Aad:2014aqa}. For
$f=1000$~GeV, we find that $c_{sg}$ values up to about 100 (which
corresponds to an effective coupling of about $10^{-3}$) are allowed
independently of the other parameters, and we adopt this upper limit
henceforth.

Turning our attention to the dark matter phenomenology, we first study the
$\eta$ relic abundance $\Omega h^2\rvert_{\eta}$. This is numerically
computed in Section~\ref{sec:mdmonojets_lhc} with the {\sc Mi\-crO\-me\-gas}
package~\cite{Belanger:2008sj}, in which we have implemented our model via
{\sc FeynRules}~\cite{Alloul:2013bka}. An approximate expression describing
 the relevant total thermally-averaged annihilation cross
section $\left\langle\sigma v\right\rangle$ can nonetheless
be derived analytically. Restricting ourselves to the leading $S$-wave
contribution and ignoring all possible special kinematic configurations
featuring, \eg, intermediate resonances, the thermally-averaged cross
section associated with $\eta$ annihilation into a pair of gluons is given by
\begin{align}
\label{eq:mdmonojets_RDgg}
\left\langle \sigma v \right\rangle_{gg} \simeq 
\frac{\alpha_s^2 c_{sg}^2 m_\eta^2 \left( c_{s\eta} f^2 + 4 c_{\partial s\eta} m_\eta^2 \right)^2}{16\pi^3 f^4 \left( m_s^2 - 4 m_\eta^2 \right)^2}~. 
\end{align}
In the case where $m_\eta > m_s$, an additional $2\to 2$ annihilation channel
contributes, $\eta\eta \to s s$, whose leading $S$-wave contribution reads
\begin{align}\label{eq:mdmonojets_RDss}
\left\langle \sigma v \right\rangle_{ss} \simeq 
\frac{\sqrt{1 - \frac{m_s^2}{m_\eta^2}} \left( c_{\partial s\eta} m_s^2 + c_{s\eta} f^2 \right)^4}{16 \pi f^4 m_\eta^2 \left( m_s^2 - 2 m_\eta^2 \right)^2}\ .
\end{align}
We impose the requirement that the $\eta$ relic density satisfies the upper
limit~\cite{Ade:2015xua}
\be
  \Omega h^2\rvert_{\eta} \leq \Omega h^2\rvert_\textrm{exp}
     = 0.1188\pm 0.0010\, .
\label{eq:OMh}\ee
Assuming a standard thermal freeze-out mechanism, and ignoring singular parameter space regions such as resonances, 
the dark matter relic density does not depend strongly on whether $m_\eta > m_s/2$ or $< m_s/2$. This condition is, 
however, crucial for the LHC: monojet searches can typically only reach couplings that correspond to thermal self-annihilation
cross sections once the mediator can be produced and decay on-shell. Instead, in the off-shell regime, the LHC tends
to probe parameter space regions where the dark matter abundance lies below
$\Omega h^2\rvert_\textrm{exp}$~\cite{Feng:2005gj}, but there are exceptions~\cite{Busoni:2014gta}. 
Finally, regardless of the momentum-dependent or -independent nature of the dark matter
interactions the dominant contribution to the dark matter annihilation comes from the $S$-wave.

Direct detection searches yield additional constraints on the phenomenologically
viable regions of the model parameter space. These however do not constrain the strength of the momentum-dependent interactions, as the corresponding scattering cross section is proportional to the dark matter-nucleus momentum transfer which is very small compared to the mediator mass. On the other hand, the
momentum-independent couplings in Eq.~\eqref{eq:mdmonojets_thelagrangian} lead
to an effective interaction between $\eta$ particles and gluons,
\begin{equation}
\label{eq:mdmonojets_DDlaggluons}
{\cal{L}}_{\eta g} = f_G ~ \eta^2 ~ G_{\mu\nu}G^{\mu\nu}
\qquad \text{with}\qquad
f_G = \frac{\alpha_s c_{sg} c_{s\eta}}{32 \pi} \frac{1}{m_s^2} \ .
\end{equation}
The spin-independent dark matter scattering cross section $\sigma_{\rm SI}$ is
then found to take the form~\cite{Hisano:2010ct,Chu:2012qy}
\begin{equation}
\sigma_{\rm SI} = \frac{1}{\pi}\
    \bigg(\frac{m_\eta m_p}{m_\eta + m_p}\bigg)^2\
    \left| \frac{8\pi}{9 \alpha_s} \frac{m_p}{m_\eta} f_G f_{TG} \right|^2\ ,
\label{eq:sigSI}\end{equation}
where the term inside the brackets corresponds to the DM-nucleon reduced mass,
and the squared matrix element depends on the gluon form factor $f_{TG}$. The
latter is derived from the quark form factors $f_{Tq}$~\cite{Shifman:1978zn},
\be
  f_{TG} = 1 - \sum_{q=u,d,s} f_{Tq}\ ,
\ee
for which we take the values \mbox{$f_{Tu}=0.0153$}, \mbox{$f_{Td}=0.0191$} and
\mbox{$f_{Ts}=0.0447$}~\cite{Belanger:2014vza}. The above expression for 
$f_{TG}$  would change if additional couplings between the mediator $s$ and quarks were introduced. Our predictions for $\sigma_{\rm SI}$
are compared, in the next section, to limits extracted from LUX
data~\cite{Akerib:2015rjg}%
\footnote{While this work was being completed, the LUX collaboration has updated
their results on the basis of 332 live days of exposure~\cite{Akerib:2016vxi}.
We do not include the latest limits in our analysis. Although more constraining,
the new LUX results do not imply significant differences in the allowed region
of the parameter space.}.

Finally, additional restrictions can also be imposed on the model from
perturbative unitarity requirements. For a given process, the effective
Lagrangian in Eq.~\eqref{eq:mdmonojets_thelagrangian} is indeed expected
to provide an accurate description of the underlying physics only as long as the
typical momentum involved lies below a cutoff scale which we have so far
kept unspecified. This scale can be deduced rigorously on a model-by-model
basis, but its minimal acceptable value can be estimated without referring to
any
specific ultraviolet completion. By enforcing the $S$ matrix to be
perturbatively unitary, we ensure that calculations performed on the basis of
the Lagrangian of Eq.~\eqref{eq:mdmonojets_thelagrangian} provide reliable
predictions~\cite{Shoemaker:2011vi,Endo:2014mja,Kahlhoefer:2015bea}.

We provide the details of the calculation in Appendix~\ref{pertunitarity}, where we show that 
perturbative unitarity of the $gg \rightarrow \eta\eta$ scattering amplitude imposes the constraints
\begin{equation}\label{eq}
\kappa_{\rm MI} < \frac{64 \sqrt{2} \pi^2 (1 - \frac{m_s^2}{Q^2})}{\alpha_s \left( 1 - \frac{4 m_\eta^2}{Q^2}  \right)^{1/4}}\ ,
\end{equation}
where \mbox{$\kappa_{\rm MI} = c_{s\eta} c_{sg}$} and
\begin{equation}
\label{eq:KMD}
\kappa_{\rm MD} < \frac{64 \sqrt{2} \pi^2 f^2 (Q^2 - m_s^2)}{\alpha_s Q^4 \left( 1 - \frac{4 m_\eta^2}{Q^2}  \right)^{1/4}}\ ,
\end{equation}
with \mbox{$\kappa_{\rm MD} = c_{\partial s\eta} c_{sg}$}. In typical hadron
collider processes like those occuring at the LHC, the scale $Q^2$ varies from
one event to another. In order to simplify the discussion, we judiciously focus
on large values of $Q^2$ that are relevant for the high-energy tail of the
differential distributions where the effective theory is expected to break down.
Considering typical missing transverse-momentum distributions related to
monojet events and the current LHC luminosity, the tail of the distribution
extends to $|Q|\sim 2$ TeV while most events relevant for the
extraction of LHC constraints feature a missing transverse energy in the
[700, 1500]~GeV range. In the next section, unitarity bounds are therefore computed
for the conservative choice $|Q|= 2$~TeV.

%% file: 3-results.tex
\label{sec:mdmonojets_lhc}

We now estimate the constraining power of monojet
searches both in the case of momentum-dependent and momentum-independent
interactions. For simplicity, we consider scenarios featuring either momentum-independent ($c_{\partial s\eta} = 0$) or momentum-dependent ($c_{ s\eta} = 0$) couplings, and we set the composite scale $f$ to 1~TeV. The mediator coupling to the gluon field
strength tensor is fixed to $c_{sg} =$ 10 and 100, as allowed by the dijet bounds discussed
in Section~\ref{sec:mdmonojets_ourmodel}. We finally discuss the complementarity
between theory, collider and cosmological constraints.

\subsection{Analysis setup}\label{sec:mdmonojets_setup}

In order to evaluate the LHC sensitivity to our model via monojet probes, we
compare our theoretical predictions to official ATLAS results based on early
13~TeV data at an integrated luminosity of 3.2 fb$^{-1}$~\cite{Aaboud:2016tnv}.
This is achieved via an implementation of
the analysis of Ref.~\cite{Aaboud:2016tnv} in
the {\sc MadAnalysis 5} framework~\cite{Conte:2012fm, Conte:2014zja}. Details on
our code and its validation are publicly available on
{\sc Inspire}~\cite{13tevmonojet} and within the {\sc MadAnalysis 5} Public
Analysis Database~\cite{Dumont:2014tja}
\footnote{\url{https://madanalysis.irmp.ucl.ac.be/wiki/PublicAnalysisDatabase}.}.
Our recasted analysis is in agreement with the ATLAS official
results for well-defined event samples at the 5\% level, and
we have also compared, for consistency, our results to those obtained when using LHC
Run~I data~\cite{Brooijmans:2016vro}.

The analysis under consideration preselects events featuring one final-state
hard jet with a transverse-momentum $p_T$ larger than 250~GeV and a
pseudorapidity satisfying \mbox{$|\eta|<2.4$}, as well as at most four jets with
\mbox{$p_T> 30$~GeV} and \mbox{$|\eta|<2.8$}. Moreover,
each jet is required to be azimuthally separated from the missing
momentum by an angle \mbox{$\Delta\phi>0.4$}, and
events exhibiting muons or electrons with a transverse momentum greater than 10~GeV and 20~GeV 
respectively are vetoed. Preselected events are then
categorized into seven inclusive and six exclusive signal regions.
The seven inclusive regions are defined by seven
overlapping selections on the missing transverse energy, demanded to be
larger than 250, 300, 350, 400, 500, 600 and 700~GeV. The same thresholds are
further used to define six missing energy bins [250,300],
[300,350], [350,400], [400,500], [500,600] and [600,700]~GeV, which form the six exclusive signal regions.

In order to perform our study,
 we have implemented the model described in
Section~\ref{sec:mdmonojets_ourmodel}
in the {\sc FeynRules}~\cite{Alloul:2013bka} package, and
generated a UFO library~\cite{Degrande:2011ua} that we have imported into
\amc~\cite{Alwall:2014hca}. Hard-scattering events describing the
$pp\to\eta\eta j$ process (with an 80~GeV selection threshold on the jet $p_T$)
have been generated for a center-of-mass energy of 13~TeV and matched to
the parton showering and hadronization infrastructure of
{\sc Pythia~6}~\cite{Sjostrand:2006za}. The events are then processed by
{\sc Delphes~3}~\cite{deFavereau:2013fsa} for a fast detector simulation using
a tuned parameterization of the ATLAS detector and the
{\sc Fastjet} program~\cite{Cacciari:2011ma} for jet reconstruction by means of
the anti-$k_T$ algorithm~\cite{Cacciari:2008gp} with an $R$-parameter set to
0.4. Finally, the {\sc MadAnalysis 5} program is used to handle the event
selection and to compute the associated upper limit at the 95\% confidence level
(CL) on the signal cross section according to the CLs
technique~\cite{Read:2000ru,Read:2002hq}. Although the considered analysis
contains 13 signal regions, the upper bound on the cross section (interpreted at leading order) is determined
only from the region that is expected to be the most sensitive. This region is determined using the
background rate, its uncertainty and the observed number of events reported
by the ATLAS collaboration.

For discrete choices of the mediator mass $m_s=50$, 250, 500 and 750~GeV, we scan
over various ranges of the dark matter mass with $2m_\eta>m_s$. Since
only one of the $c_{s\eta}$ or $c_{\partial s\eta}$ parameters are taken to be non-zero at a time,
the computed cross section upper limits only depend on the kinematics of
the events and not on the overall rate. The results are thus independent of the
actual values of the $c_{s\eta}$, $c_{\partial s\eta}$ and $c_{sg}$ parameters,
and we have consequently fixed ($c_{s\eta}$,
$c_{\partial s\eta}$) to the nominal values (1, 0) and
(0, 1). This choice enables an easy rescaling of the monojet cross section when
different values of the input parameters are chosen and a straightforward
derivation of limits on the momentum-dependent and momentum-independent
interactions for a given set of masses and couplings.

In addition, we have also evaluated the LHC sensitivity to our model for a
luminosity
of 300~fb$^{-1}$, this time using {\sc Pythia 8}~\cite{Sjostrand:2014zea} for
the parton showering and hadronization of the signal samples, along with efficiency factors and smearing functions aimed at reproducing the
performance of the ATLAS detector during the first run of the
LHC~\cite{Aad:2009wy}. Thanks to the higher statistics and the different 
shape of the missing energy distribution for signal and background, 
the optimal sensitivity to the signal is expected for tighter missing 
energy requirements than those adopted in Ref.~\cite{Aaboud:2016tnv},
thus motivating extending the number of signal regions. 
This, however, requires the extrapolation of the predictions for the expected Standard Model background and the associated uncertainties.

The 3.2~fb$^{-1}$ monojet publication of ATLAS
provides the Standard Model expectation  for the missing transverse-energy 
distribution~\cite{Aaboud:2016tnv}, so that the latter can be used to
extract the expected number of background events $N_{\rm bg}$ for
300~fb$^{-1}$. The estimation of the systematic uncertainties
$\Delta N_{\rm bg}$
is however luminosity-dependent due to an extrapolation of the dominant $Z$+jets
and $W$+jets backgrounds from the number of data 
events observed in appropriate control regions to the signal
regions. We consequently parametrise $\Delta N_{\rm bg}$ as
\be
\Delta N_{\rm bg}^2=\Big(k_1\ \sqrt{N_{\rm bg}}\Big)^2 + 
    \Big(k_2\ N_{\rm bg}\Big)^2\ .
\ee
The first term on the right-hand side represents the
statistical error on the number of events observed in the control regions
and is controlled by the $k_1$ parameter, while the second term
consists of the systematic uncertainties connected to the extrapolation procedure
from the control region to the signal regions and is driven by the $k_2$
parameter. The ATLAS analysis finds the latter
to be slowly varying with the missing transverse-energy selection and
is of the order of a few percent~\cite{Aaboud:2016tnv}. We adopt the choice of
\mbox{$k_1=1.51$} and \mbox{$k_2=0.043$}, which parametrize the ATLAS results of
Ref.~\cite{Aaboud:2016tnv} at the percent level, and calculate 95\% CL upper limits
on the signal cross-section for overlapping signal regions 
defined by minimum requirements on the missing transverse
energy varying in steps of 100 GeV between 500 and 1400 GeV. The
statistical procedure relies on
a Poisson modelling with Gaussian constraints using the CLs prescription and
the asymptotic calculator implemented in the {\sc RooStat}
package~\cite{Moneta:2010pm}. The lowest upper limit on the fiducial production
cross section (with a constraint on the jet transverse momentum of
\mbox{$p_T>80$~GeV}) is then taken to be the final result.

\subsection{Bounds derived from LHC monojet data}
\label{sec:mdmonojets_cslimits}

\begin{figure}
 \begin{center}
  \includegraphics[width=0.45\textwidth]{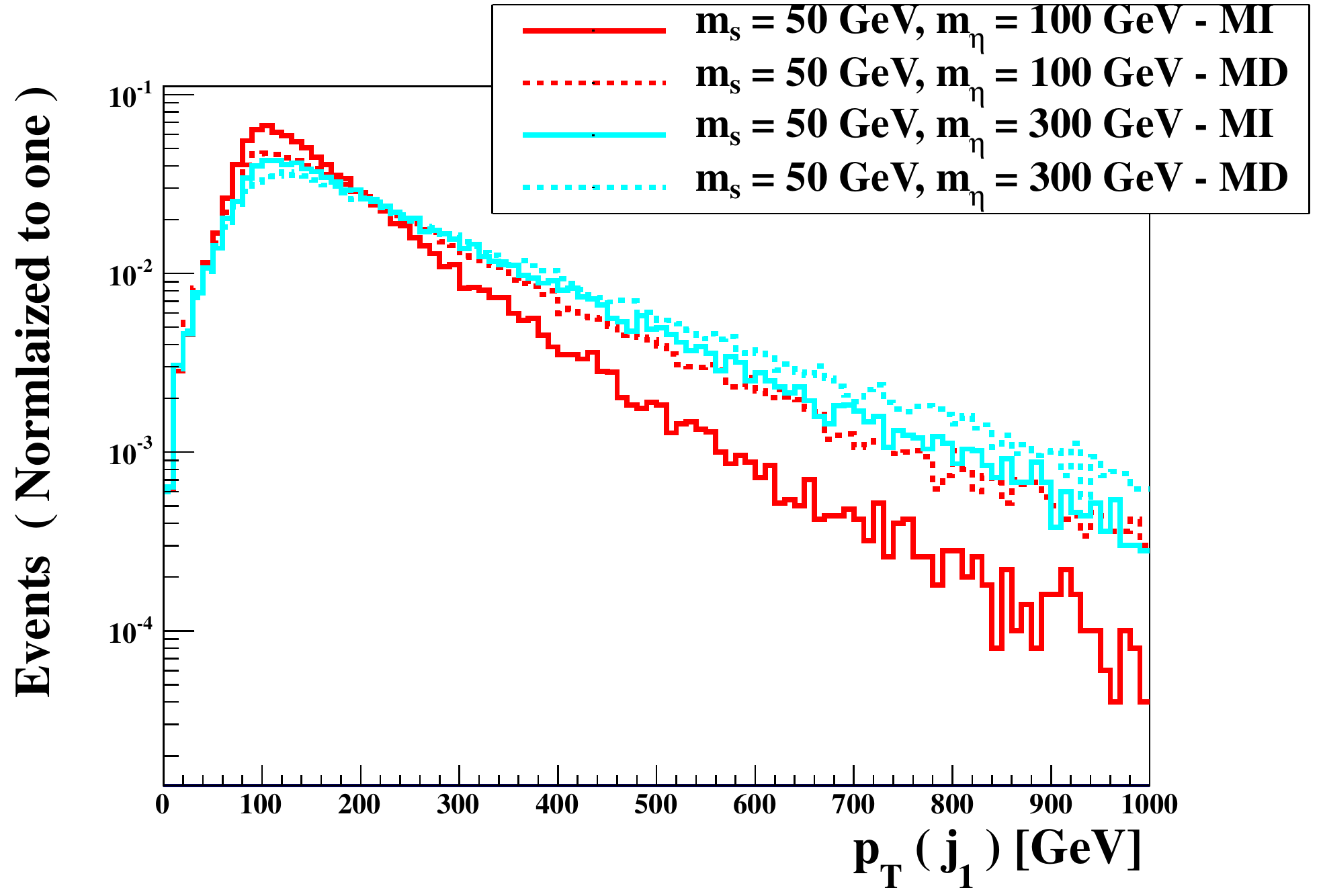}
  \includegraphics[width=0.45\textwidth]{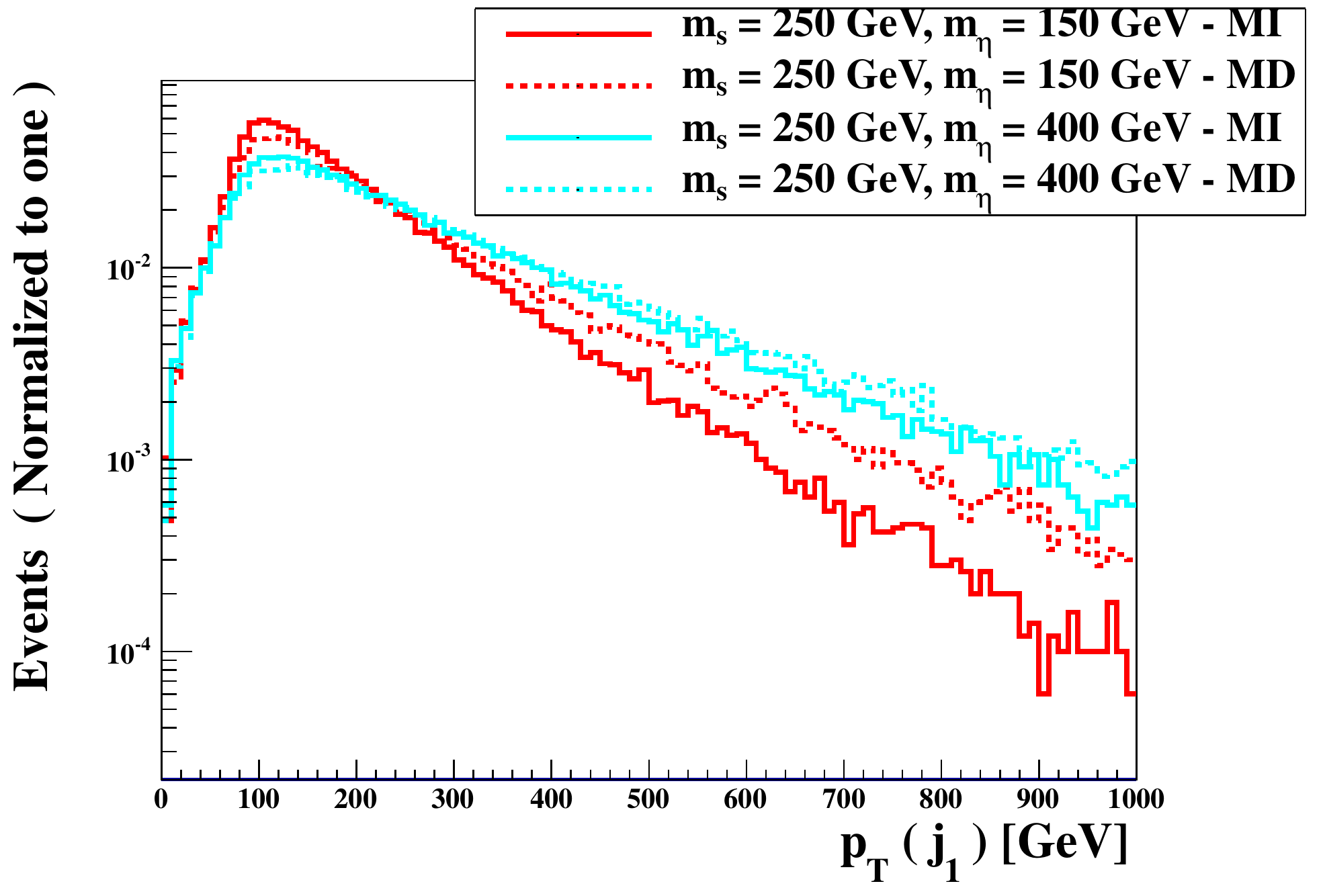}
  \caption{Normalized distributions in the transverse momentum of the leading
    jet assuming a perfect detector. We consider a mediator mass of
    50~GeV and 250~GeV in the left and right panels respectively, and a dark
    matter mass of 100~GeV and 300~GeV (left panel) or 150~GeV and 400~GeV
    (right panel). The solid lines reflect scenarios featuring
    momentum-independent (MI) interactions while the dashed lines correspond to
    scenarios featuring momentum-dependent (MD) interactions.}
  \label{fig:mdmonojets_monopt}
 \end{center}
\end{figure}

As a first illustration of the differences between scenarios featuring
momentum-independent and momentum-dependent interactions, we show
the leading jet $p_T$ distributions obtained with {\sc MadAnalysis 5}
for the representative
mass combinations \mbox{$(m_{s}, m_\eta) = (50, 100/300)$} GeV and
\mbox{$(m_{s}, m_\eta) = (250, 150/400)$~GeV} in the left and right panels of
Figure~\ref{fig:mdmonojets_monopt} respectively. Focusing on the shapes of the
distributions that have been normalized to one, one observes that momentum-dependent interactions induce a harder jet $p_T$ spectrum. As a result, one expects that a larger
fraction of events would pass a monojet selection when momentum-dependent
interactions are present.
For instance, choosing \mbox{$c_{sg} = 100$}, \mbox{$f =1$~TeV} and
either \mbox{$c_{\partial s\eta} = 2.5$} in the momentum-dependent case or
\mbox{$c_{s\eta}= 0.5$} in the momentum-independent case we obtain, in both scenarios, a fiducial
cross section of 2.9~pb once an 80~GeV generator-level selection on the leading
jet $p_T$ is enforced. The efficiency associated with a transverse-momentum
selection of \mbox{$p_T > 300$~GeV} on the leading jet is, however,
relatively larger by about 50\% in the momentum-dependent case.
The difference between
the two scenarios is significantly reduced for larger dark matter
masses.

\begin{figure}
 \begin{center}
 \includegraphics[width=0.4\textwidth]{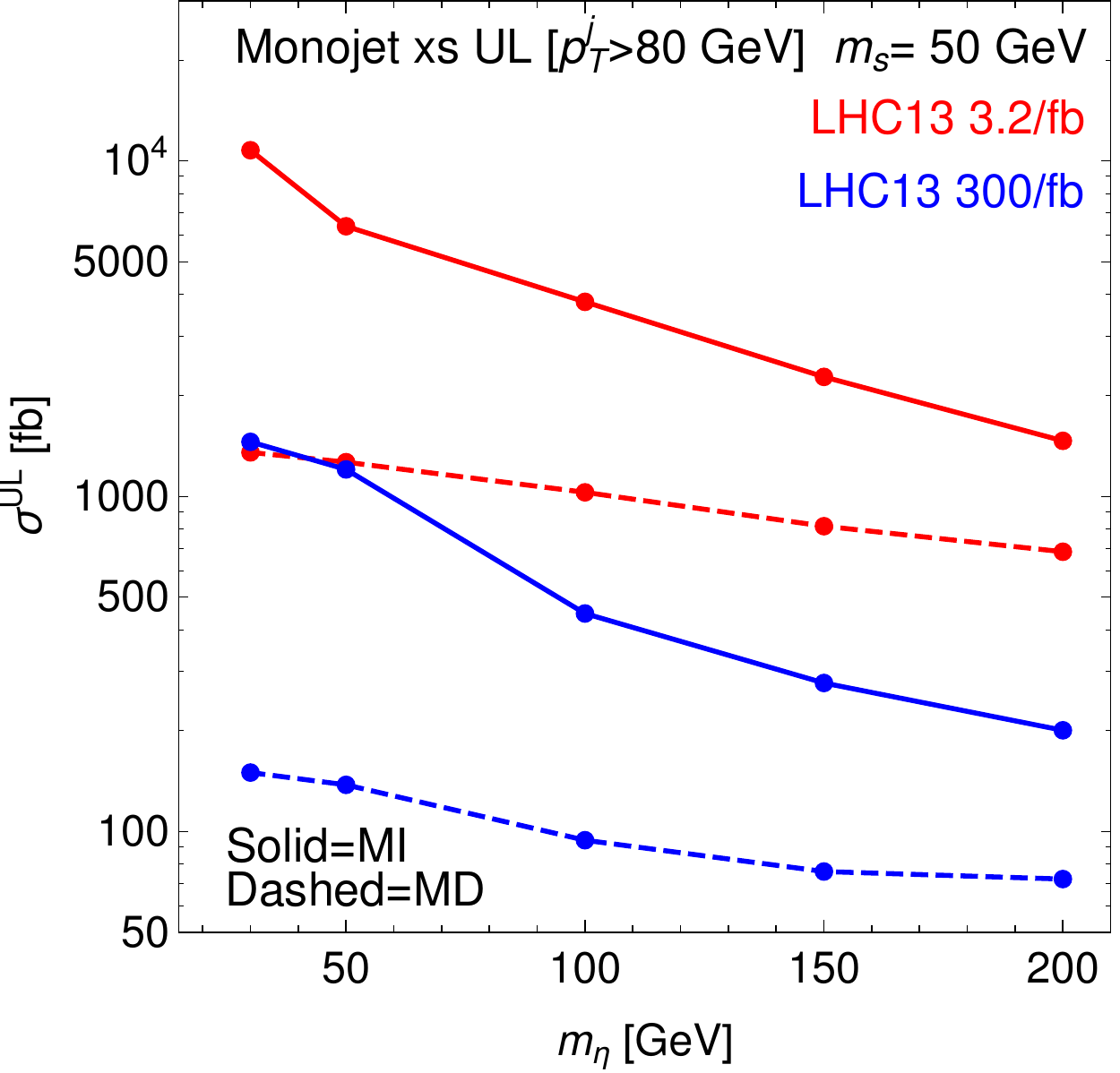}
 \includegraphics[width=0.4\textwidth]{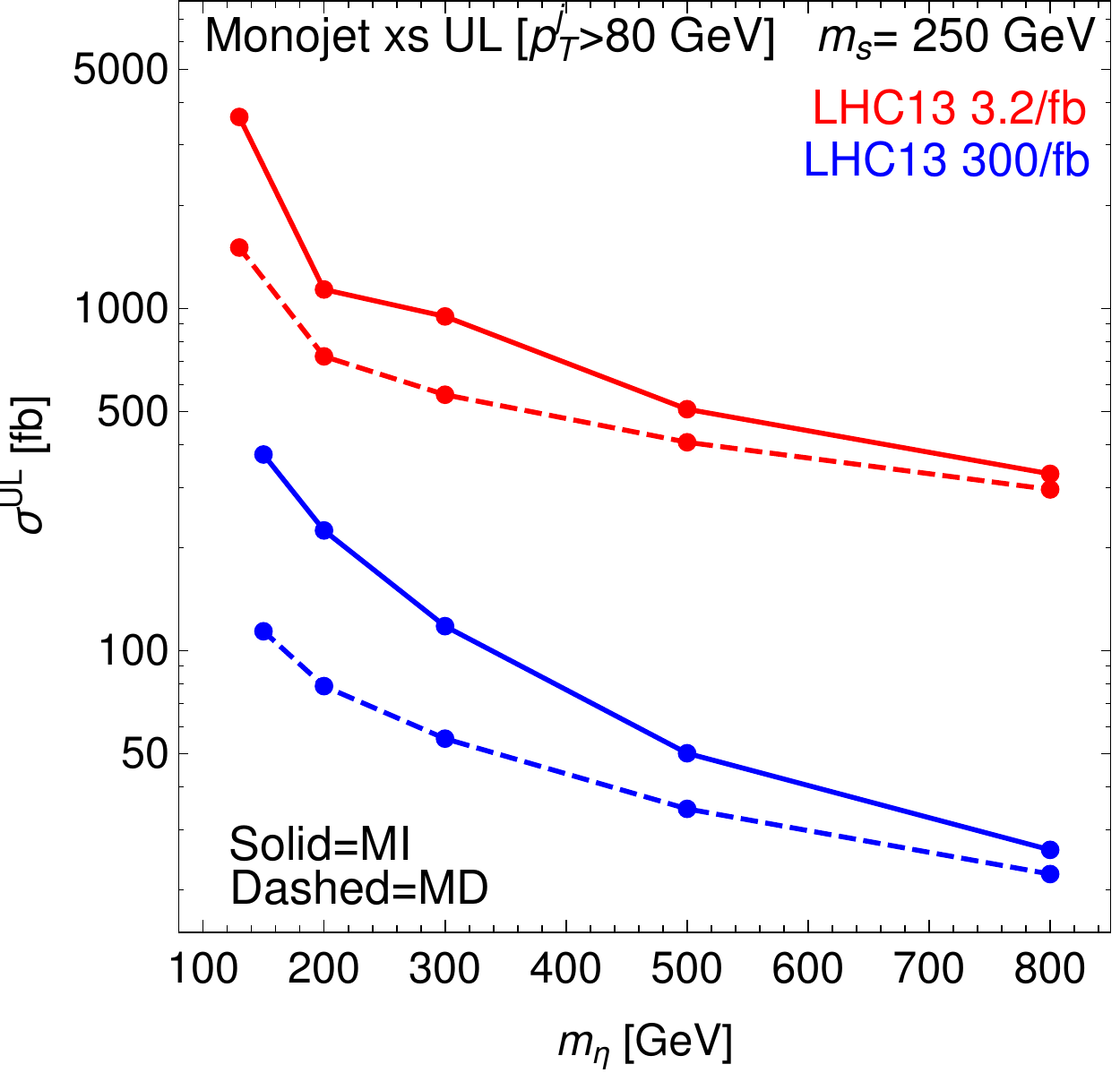}\\[.3cm]
 \includegraphics[width=0.4\textwidth]{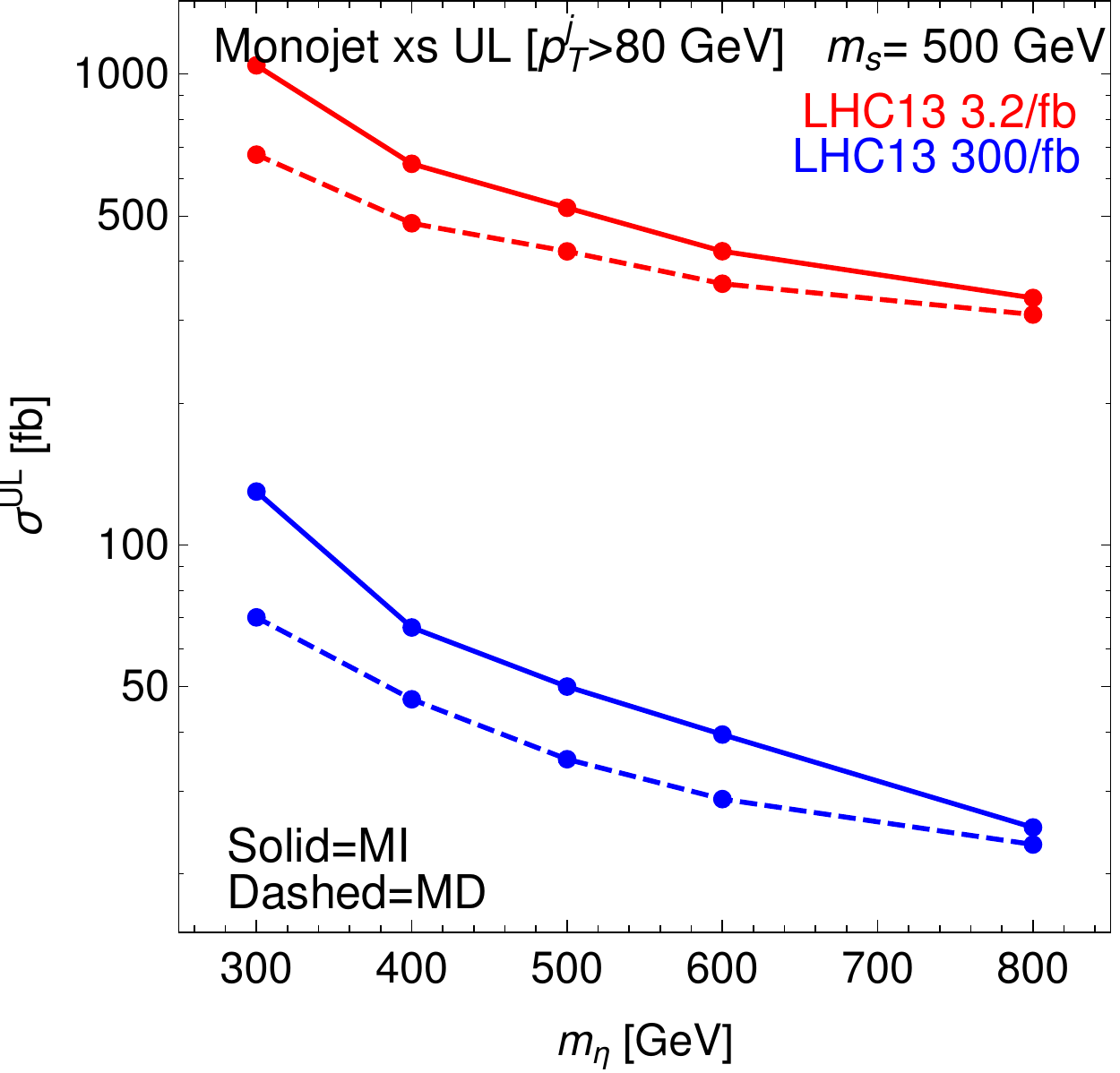}
 \includegraphics[width=0.4\textwidth]{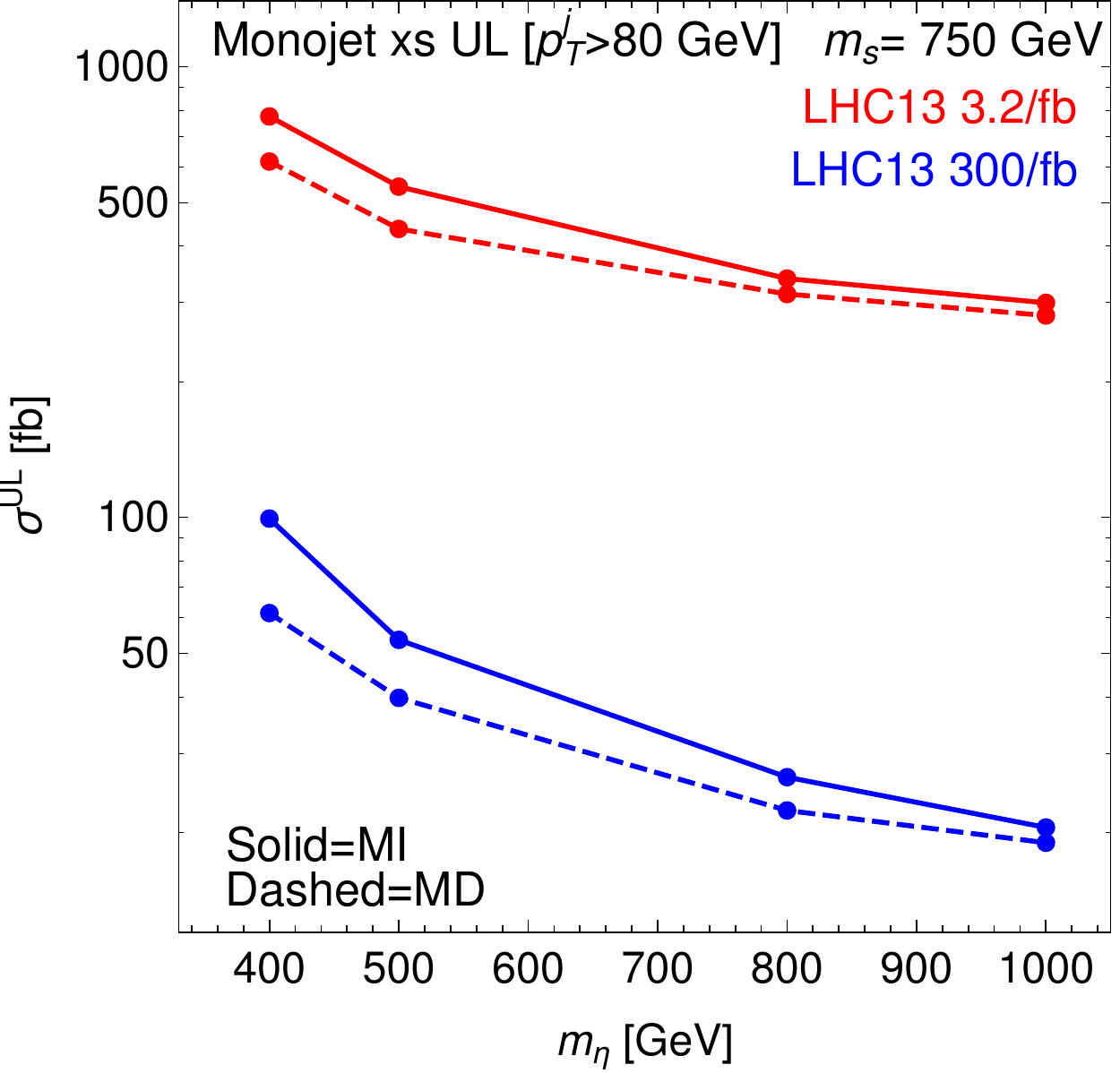}
  \caption{95\% CL upper limits (UL) on the monojet production fiducial cross
    section (that includes a generator-level selection of $p_T > 80$~GeV on the
    leading jet). We consider proton-proton collisions at a center-of-mass
    energy of 13~TeV with an integrated luminosity of 3.2~fb$^{-1}$ (recasting,
    red lines) and 300~fb$^{-1}$ (projections, blue lines) for
    \mbox{$m_s=50$~GeV} (top left), 250~GeV (top right), 500~GeV (bottom left)
    and 750~GeV (bottom right) as a function of $m_\eta$. The solid lines
    correspond to the momentum-independent case, whereas the dashed lines
    correspond to the momentum-dependent case.
   \label{fig:mdmonojets_ulimits}}
\end{center}
\end{figure}

As explained in Section~\ref{sec:mdmonojets_setup}, for a given value of $m_s$ the constraints that can be
derived from LHC dark matter searches only depend on $m_\eta$, since in the
relevant subprocesses the mediator
has to be off-shell. In Figure~\ref{fig:mdmonojets_ulimits}, we present the
upper limits on the monojet cross section at the LHC,
$\sigma^{\rm UL}(p p \to \eta \eta j)$,
with a generator-level selection on the transverse momentum of the leading jet
of $p_T>80$~GeV. Existing constraints extracted from 3.2~fb$^{-1}$ of 13~TeV
LHC collisions are depicted by red lines for the momentum-independent (solid)
and dependent (dashed) cases. As anticipated, the cross sections excluded at the
95\% CL are significantly smaller in the momentum-dependent setup than in the
moment-independent one, so that the former is more efficiently constrained than
the latter. We additionally
observe that the exclusion bounds become stronger with
increasing $m_\eta$. As long as enough phase space is available, larger $\eta$
masses imply a larger amount of missing energy so that the signal regions of the
monojet analysis are more populated and stronger limits can be
derived, as shown in the figure.

Our results confirm the findings of Figure~\ref{fig:mdmonojets_monopt},
the differences between the momentum-independent and
momentum-dependent cases being maximal for small values of $m_\eta$. Eventually,
for dark matter masses of about 1~TeV, the limits become identical for
both cases although the LHC loses its sensitivity for such heavy dark matter scenarios.

We also report in Figure~\ref{fig:mdmonojets_ulimits} projections for
300~fb$^{-1}$ of LHC collisions at a center-of-mass energy of 13~TeV. The
blue solid and dashed lines respectively represent the momentum-independent and
momentum-dependent cases. We observe a behavior that is similar to the lower
luminosity one, although it is now driven by the additional higher
missing-energy requirements. In the relevant bins, the signal acceptance is
again found to be higher for the momentum-dependent dark matter coupling case, so that the
corresponding exclusion bounds are stronger.
Moreover, the two classes of dark matter operators can
still only be distinguished up to a given dark matter mass, which is nonetheless larger than for lower luminosities.

\begin{figure}
 \begin{center}
  \includegraphics[width=0.4\textwidth]{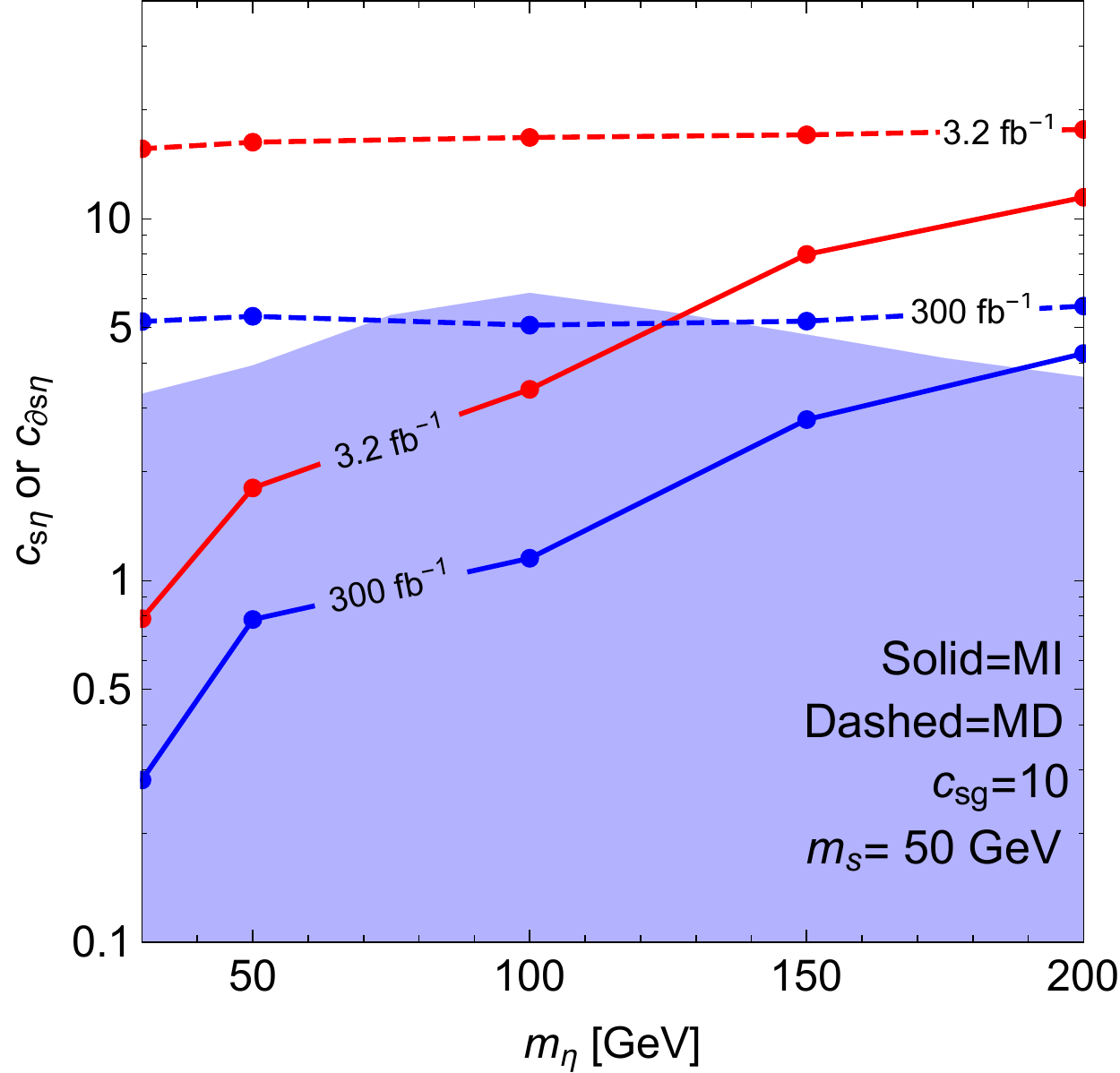}
  \includegraphics[width=0.4\textwidth]{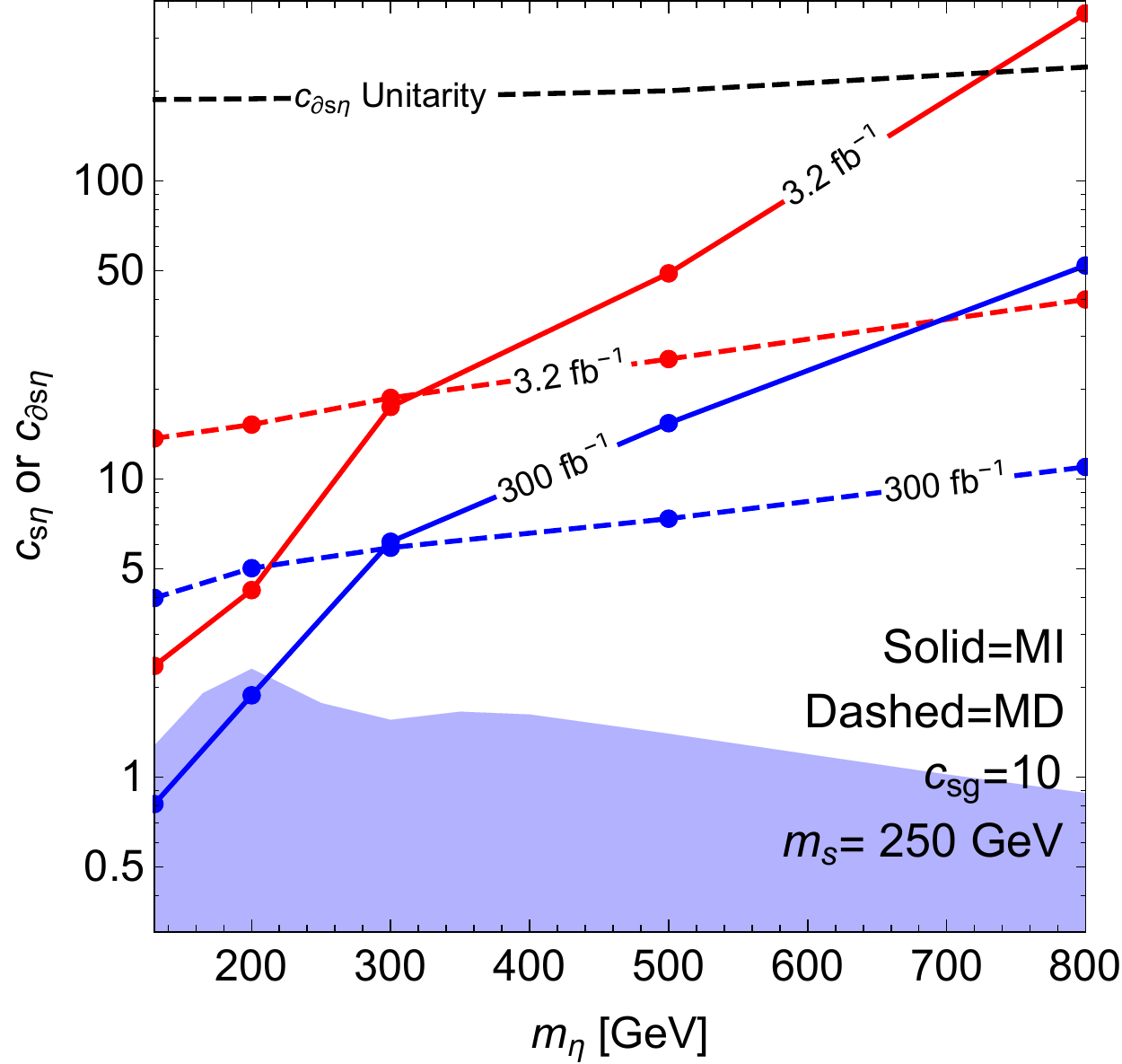}
  \includegraphics[width=0.4\textwidth]{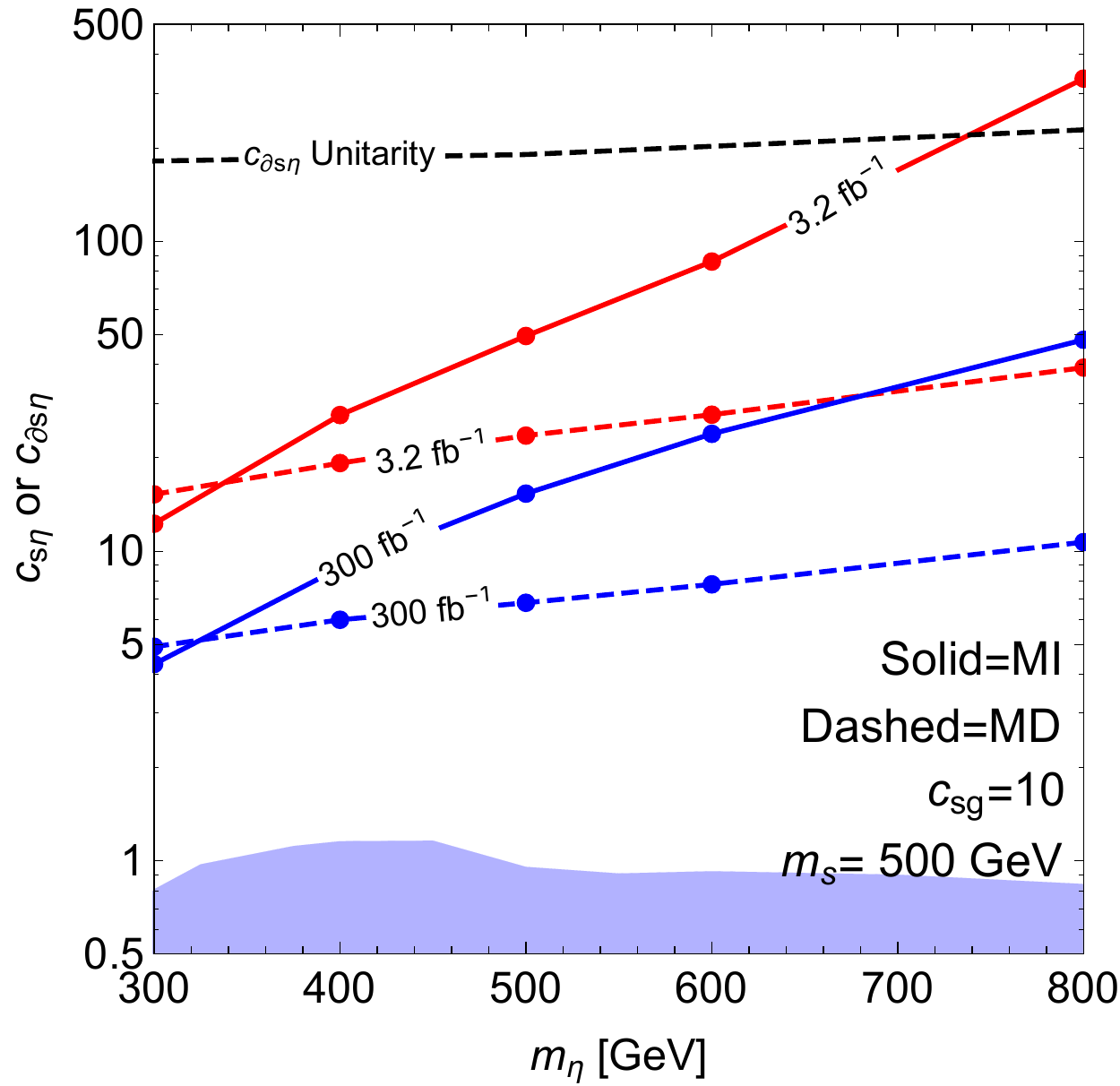}
  \includegraphics[width=0.4\textwidth]{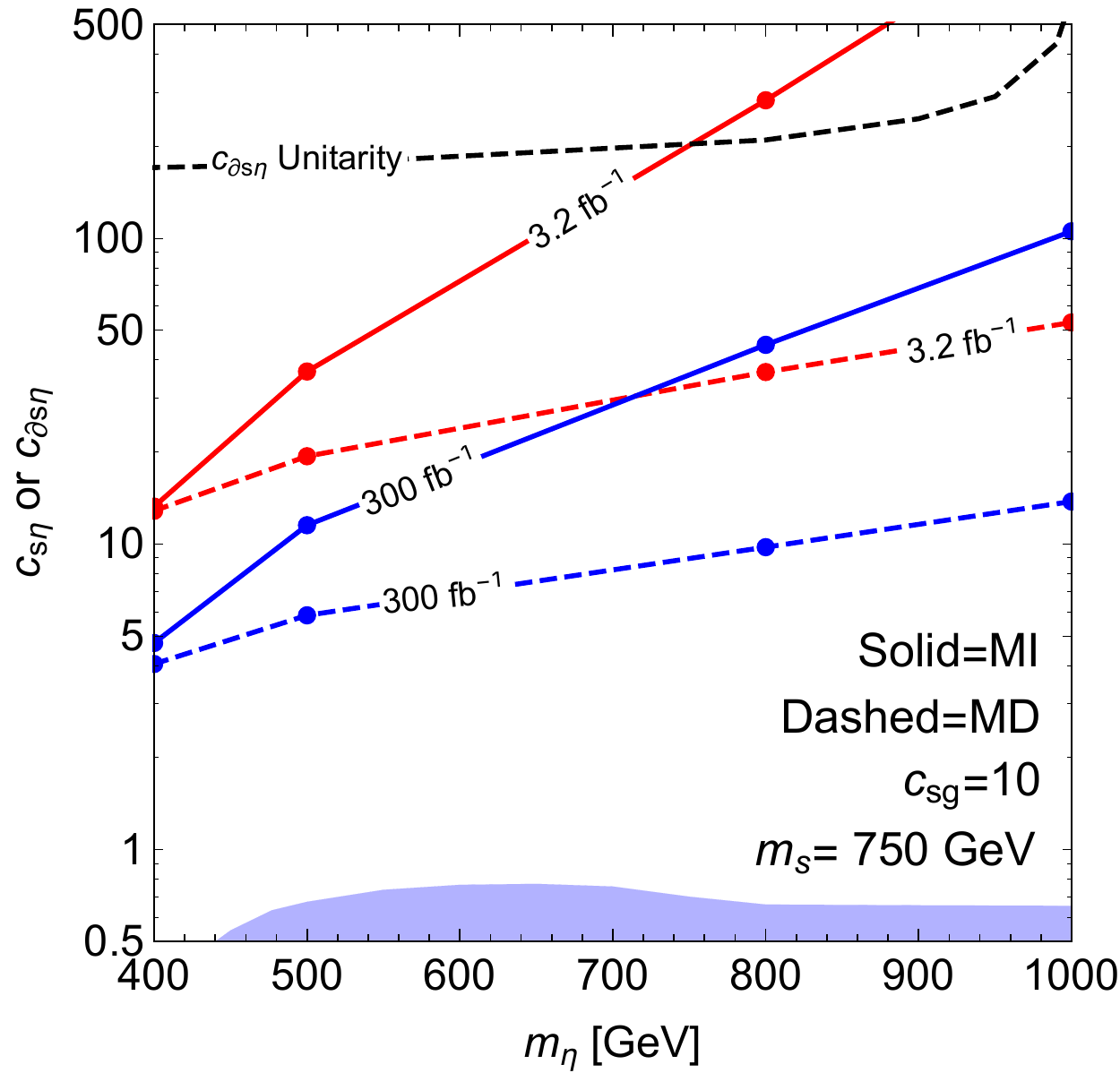}
  \caption{Constraints on the $c_i$ couplings of
     Eq.~\eqref{eq:mdmonojets_thelagrangian} driven by monojet searches. The red lines depict 
     constraints from existing 3.2 fb$^{-1}$ of data whereas blue ones correspond to predictions for an integrated
     luminosity of 300 fb$^{-1}$. We fix $f = 1$~TeV, \mbox{$m_s = 50$}~GeV (top left), 250~GeV
     (top right), 500~GeV (bottom left) and 750~GeV (bottom right) and the
     results are represented as functions of $m_\eta$ for \mbox{$c_{sg}=10$}.
     The shaded regions correspond to momentum-dependent coupling values for
     which the universe is overclosed, while above the black lines the perturbative unitarity
     of the effective theory is lost. The solid and dashed lines correspond to the momentum-independent (MI) and dependent (MD)
     cases respectively. 
     \label{fig:mdmonojets_bounds10}}
\end{center}
\end{figure}

\begin{figure}
\begin{center}
\includegraphics[width=0.4\textwidth]{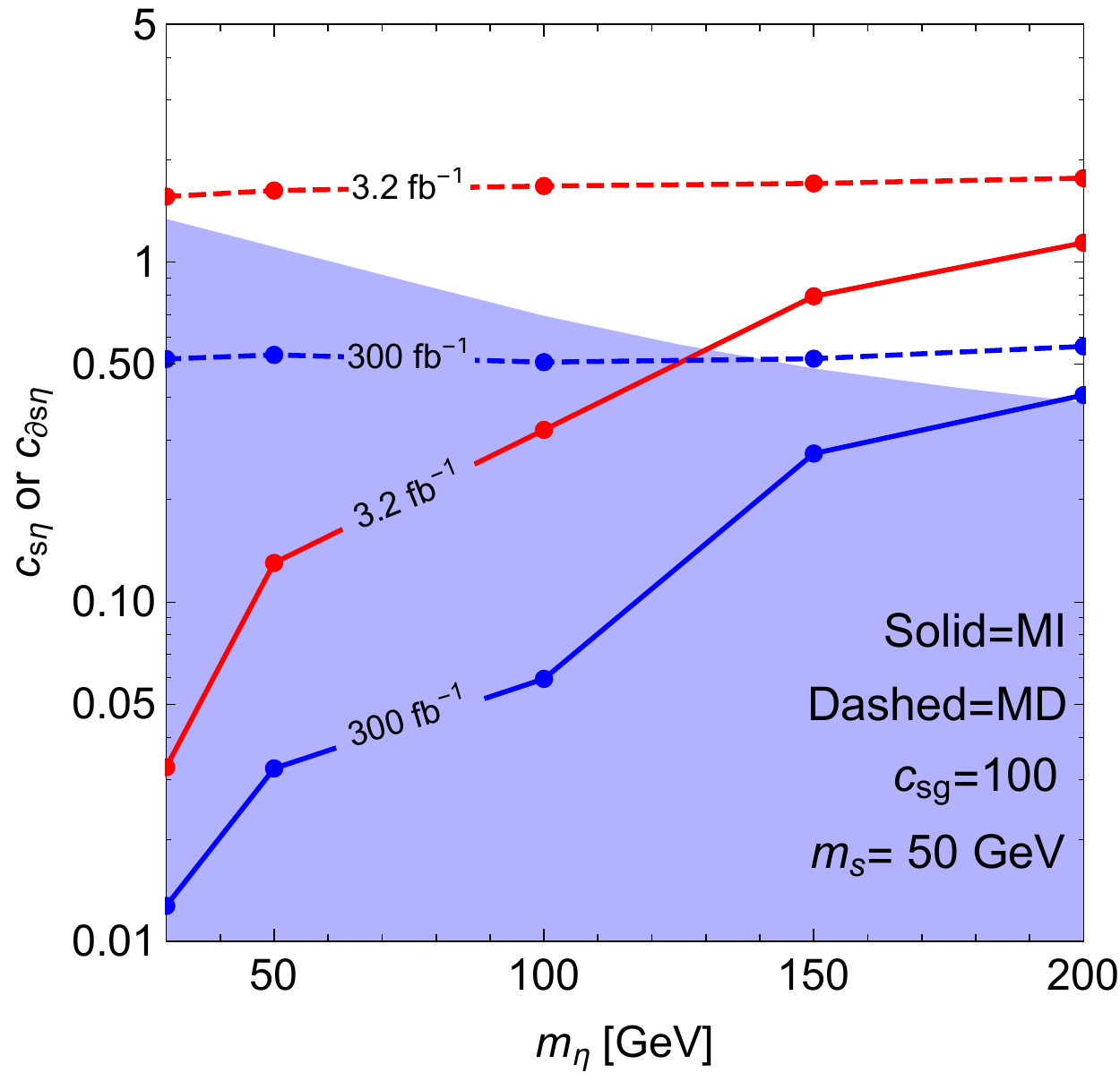}
\includegraphics[width=0.4\textwidth]{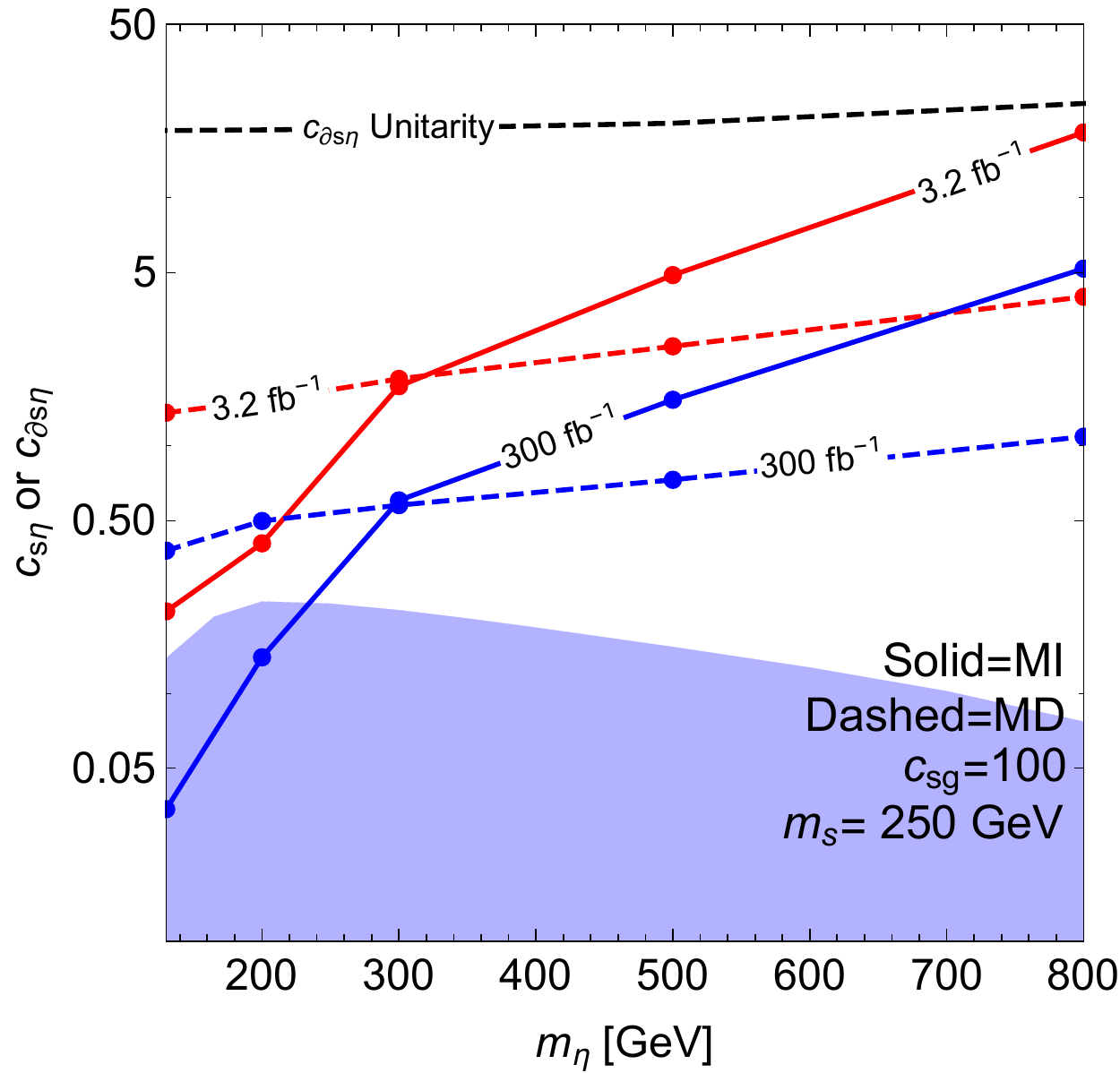}
\includegraphics[width=0.4\textwidth]{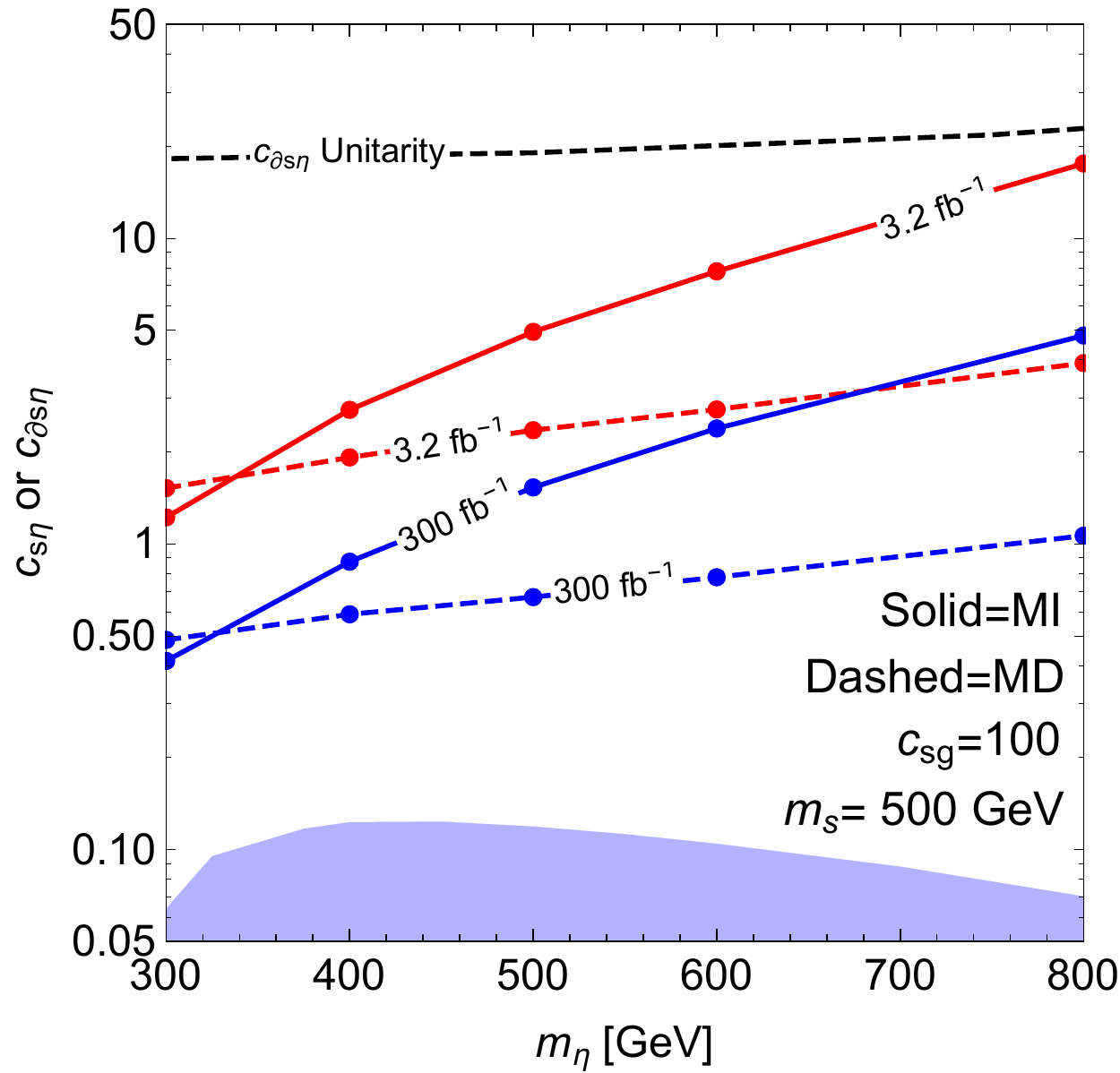}
\includegraphics[width=0.4\textwidth]{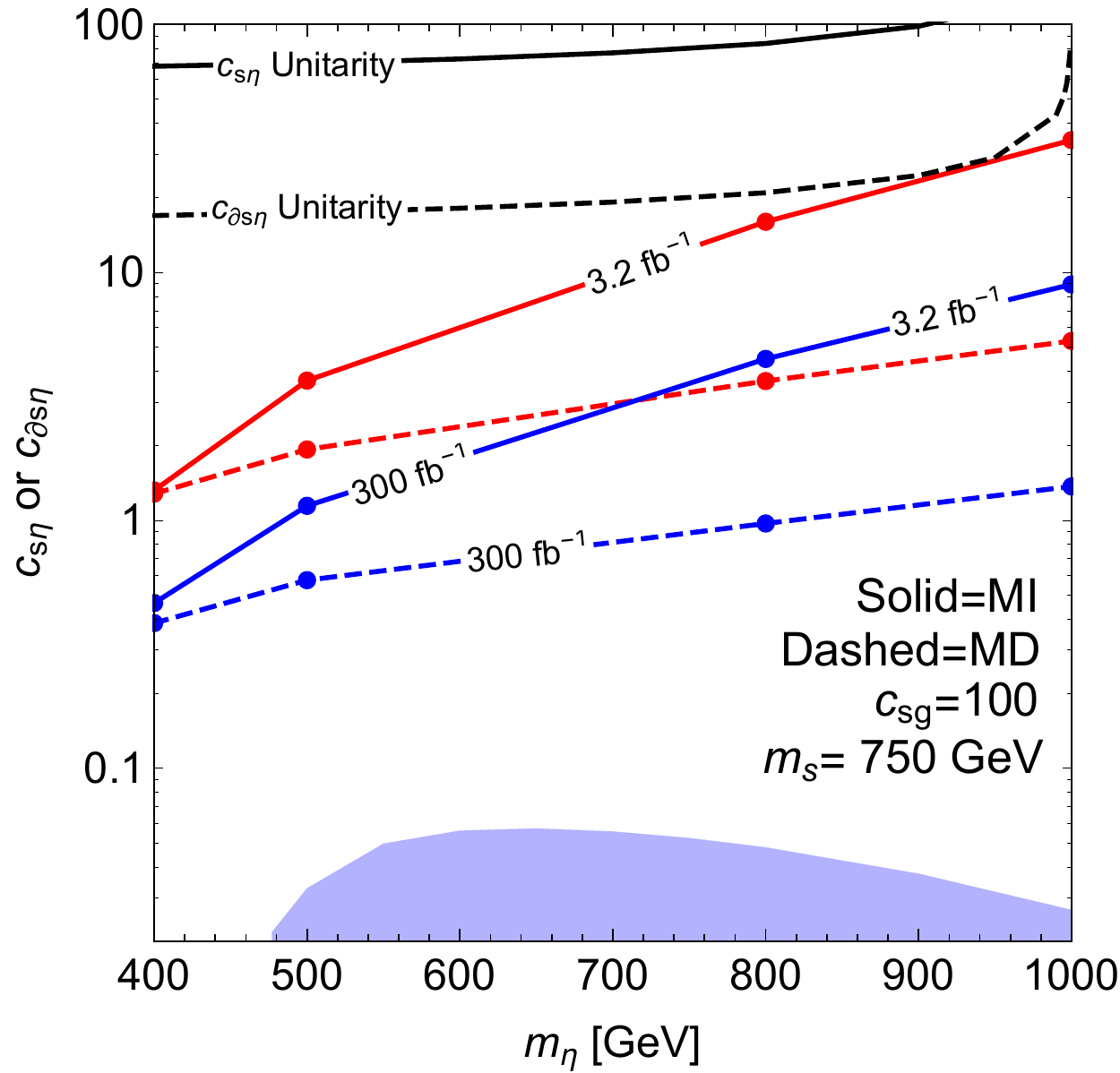}
 \caption{Same as in Figure~\ref{fig:mdmonojets_bounds10} but for $c_{sg}=100$.
  \label{fig:mdmonojets_bounds100}}
\end{center}
\end{figure}

\subsection{Complementarity of collider, cosmological and theoretical
considerations}\label{sec:compl}

In order to estimate the regions of the model parameter space that are viable with
respect to current data, we investigate the interplay between the LHC monojet bounds
presented in the previous section and the dark matter and theoretical
considerations discussed in Section~\ref{sec:mdmonojets_ourmodel}.
Assuming momentum-independent dark matter interactions, the LUX results exclude the
spin-independent direct detection cross section predicted by Eq.~\eqref{eq:sigSI} in the entire parameter
space region accessible with the 13~TeV LHC monojet results.
More precisely, for a dark matter mass of 50~GeV that is close to the LUX
sensitivity peak and for the minimal \mbox{$c_{sg} = 10$} choice, the maximum
$c_{s\eta}$ allowed values are of the order of $1.2\times 10^{-3}$, 0.03, 0.13
and 0.28 for \mbox{$m_s = 50$}, 250, 500 and 750~GeV respectively. Increasing
the dark matter mass to a slightly higher value of 200~GeV that is still within
the LHC reach, these numbers increase to 0.008, 0.2, 0.5 and 0.9 \footnote{These numbers assume that the local dark matter density is
$\rho_0 = 0.3$ GeV cm$^{-3}$.}.
Therefore, in the momentum-independent case, an observable monojet signal could be explained only by missing energy unrelated to dark matter.
Thus, in the following, we show the constraint from the dark matter relic density only for the momentum-dependent case.

In Figures~\ref{fig:mdmonojets_bounds10} and~\ref{fig:mdmonojets_bounds100} we superimpose constraints arising from
the 13~TeV LHC monojet search results and the corresponding projections (red and blue lines respectively) 
on those obtained by imposing the relic density bound of
Eq.~\eqref{eq:OMh} (the latter for the momentum-dependent case only), assuming a
standard thermal freeze out dark matter scenario. The bounds on the $c_{s\eta}$ coupling
are stronger for larger $f$ values while those on the
$c_{\partial s \eta}$ parameter are weaker.
In the shaded regions, $\eta\eta$ annihilation is not efficient enough, so that
the Universe is overclosed.
Along the borders of these regions, the relic density limit is exactly
reproduced, the shape of these borders being fairly well described by the
approximate results of Eq.~\eqref{eq:mdmonojets_RDgg} and
Eq.~\eqref{eq:mdmonojets_RDss}. In the un-shaded region, the predicted
abundance is smaller than the observed Planck value.

Since in our parameter space scan, no resonant configuration can occur, the
$c_{\partial s\eta}$ values satisfying the dark matter abundance bounds vary
relatively mildly with the dark matter and mediator masses. The minor apparent
features (especially in the $m_{s} = 250$~GeV and 500~GeV scenarios) that can
be observed are related to the opening of the additional dark
matter annihilation channel into $s$ pairs. For our choices of parameters, however,
this channel only contributes subleadingly to the relic density, its maximal impact
being found to be of the order of $15\%$. The annihilation cross section
hence approximately scales as $(c_{\partial s\eta} \times c_{sg})^2$, so that
a smaller value of $c_{sg}$ implies almost proportionally larger allowed values
for $c_{\partial s\eta}$.

We also include in the figures the perturbative unitarity limit of validity of
our effective parameterization (black dashed line)  choosing \mbox{$\left| Q \right| = 2$~TeV} in
Eq.~\eqref{eq:KMD}. For $c_{sg}=10$ the perturbative unitarity bound reads $c_{\partial s \eta} \lesssim 190$ in the momentum-dependent case,
and $c_{s \eta} \lesssim 760$ in the momentum-independent case. This bound depends weakly on $m_s$ and $m_\eta$, however it
is proportional to  $1/c_{sg}$, as observed on comparing
Figure~\ref{fig:mdmonojets_bounds10} with Figure~\ref{fig:mdmonojets_bounds100}.
We observe that although the unitarity limits do exhibit some overlap with the current and projected LHC reach,
our effective description is consistent over most of the parameter space.


Our findings show that existing monojet constraints are not yet strong enough to probe regions of parameter space where $\eta$ can account for the entire dark matter energy 
density of the Universe. We therefore recover the fairly well-known result that in the ``off-shell'' regime of dark matter models, the LHC tends to be sensitive to
dark matter candidates for which the relic density is underabundant ~\cite{Abercrombie:2015wmb,Arina:2016cqj}. On the other hand, collider searches probe large values of the $s \eta \eta$ coupling while the Planck results instead constrain small 
values, where the Universe tends to be overclosed. In this sense, there is an interesting complementarity between collider and cosmological measurements. Besides, we 
observe that for an integrated luminosity of 300 fb$^{-1}$, the LHC will be able to access a part of the low-mass region of our model where the observed dark matter 
abundance can be exactly reproduced, for dark matter masses up to about $140$ GeV. Whether or not it will be able to actually \textit{distinguish} between the two 
scenarios will be the subject of forthcoming work. Another remark is related to the fact that our results are valid regardless of the stability of the $\eta$ states 
at cosmological timescales. In other words, letting aside model-building considerations, our analysis holds for metastable $\eta$ particles as well, as long as 
they do not decay within the LHC detectors.

%% file: 4-conclusions.tex
\label{sec:conclusion}

Momentum-dependent couplings between dark matter and the Standard Model are well-motivated both from a theoretical and a phenomenological perspective. Indeed, broad classes of ultraviolet-complete dark matter models predict effective derivative operators at low energy, in particular whenever the dark matter particle is an approximate Goldstone boson of the underlying theory. This is quite natural in the context of composite Higgs models for the electroweak symmetry breaking. On the phenomenological side, scenarios involving momentum-dependent couplings can reproduce the observed dark matter abundance in the Universe, while evading the stringent bounds from direct detection experiments.

Monojet searches at the LHC are important probes of dark matter. In the context of a simplified model where a pair of dark matter particles $\eta$ interacts with the Standard Model via a scalar mediator $s$, we considered two types of dark matter-mediator couplings, momentum-dependent or momentum-independent, corresponding to two operators with different Lorentz structures. The high-energy tail of the monojet differential distribution being harder in the momentum-dependent case, the associated cross-section is expected to be more efficiently constrained by current LHC data. We demonstrated this by studying the monojet cross section upper limits in the two scenarios, employing early 13 TeV LHC data. We showed that, indeed, one can probe smaller cross sections in the momentum-dependent case. The difference in sensitivity appears when the mediator is produced off-shell, $m_\eta > m_s/2$, and provided enough phase space is available, $m_\eta \lesssim 1$ TeV.
This difference ranges from a factor of order one, for $m_\eta \sim m_s \sim 500$ GeV, up to one order of magnitude for lower masses, $m_\eta\sim m_s \sim 50$ GeV.  We moreover estimated the reach of the LHC assuming an integrated luminosity of 300 fb$^{-1}$ in proton-proton collisions at a center-of-mass energy of 13 TeV.

In the momentum-dependent case, that is free from direct detection constraints, we compared the monojet upper bounds on the dark matter couplings with the requirement of not exceeding the observed dark matter abundance in the Universe. While the present bounds correspond to under-abundant relic densities, the projected 300 fb$^{-1}$ bounds become sensitive to the observed dark matter relic density for sufficiently light masses, $m_\eta,m_s\lesssim 100$ GeV. We also carefully checked that, in the relevant parameter space of the model, our description in terms of effective operators is consistent with perturbative unitarity.

Our study indicates that, in the near future, the LHC can cover the most significant portion of the parameter space in the case of a light, off-shell mediator to the dark sector. Indeed, one can progressively close the gap between the collider upper limits on the dark matter couplings, and their values preferred by cosmological observations, assuming a standard thermal history. Were a monojet signal observed, the differences in the monojet $p_T$ distribution between the momentum-dependent and the momentum-independent couplings could provide handles on the nature of the dark matter interactions with the Standard Model. One should keep in mind that, contrary to our simplifying assumption, both types of coupling can be present, but it is likely that one provides the dominant contribution. A discrimination among the two scenarios appears feasible, once  the statistics become sufficient to analyse the shape of the distributions.

%% file: 5-unitarity.tex
\label{pertunitarity}
We follow the analysis presented in
Ref.~\cite{Endo:2014mja}, using conventions essentially coinciding with those used by Jacob and Wick in their seminal paper on the computation of
scattering amplitudes in terms of helicity eigenstates~\cite{Jacob:1959at,%
Haber:1994pe}.
In order to compute the range of validity of our effective field theory
framework, we rely on the optical theorem,
\begin{align}
  {\cal{M}}_{i \rightarrow f} - {\cal{M}}^\dagger_{f \rightarrow i} =
    -i \sum_{X} \int {\rm d}\Pi _{\rm LIPS}^{X}\ (2\pi)^4 \delta^4 (p_i - p_X)\
    {\cal{M}}_{i \rightarrow X} {\cal{M}}^\dagger_{X \rightarrow f}\ ,
\end{align}
where $X$ represents a complete set of intermediate states in the amplitudes
${\cal{M}}$ and ${\rm d}\Pi _{\rm LIPS}^{X}$ the associated Lorentz-invariant phase space measure.
This relation is exact and would hold if we could compute the amplitudes
non-perturbatively, and should also hold order-by-order in perturbation theory.
The case of interest to us is the one where $f \equiv i$, which gives
\begin{equation}
2 {\rm Im}\left[ {\cal{M}}_{i \rightarrow i} \right] = -i \sum_{X} \int {\rm d}\Pi _{\rm LIPS}^{X}\ (2\pi)^4 \delta^4 (p_i - p_X)\ \left| {\cal{M}}_{i \rightarrow X} \right|^2 \ .
\end{equation}
For $2 \to 2$ reactions and adopting the center-of-momentum reference frame, all
kinematic variables can be integrated over, except the angle $\theta$ between
the collision axis and one of the final-state particle momenta,
\begin{equation}\label{eq:opticaltheorem}
2 {\rm Im}\left[ {\cal{M}}_{i \rightarrow i} \right] = \sum_{f} \beta_f \int \frac{{\rm d} \cos\theta}{16\pi}\left| {\cal{M}}_{i \rightarrow f} \right|^2 \ ,
\end{equation}
where $\beta_f$ reads
\begin{equation}
\beta_f = \frac{\sqrt{\left[ s - (m_1 + m_2)^2 \right] \left[ s - (m_1 - m_2)^2 \right]}}{s} \ ,
\end{equation}
with $\sqrt{s}$ being the center-of-mass energy and $m_{1}$ and $m_{2}$ the masses of
the outgoing particles.
\\
\\
The scattering amplitudes ${\cal{M}}_{i \rightarrow f}$ can be expanded in
partial waves as
\begin{equation}\label{eq:expansion}
{\cal{M}}_{i \rightarrow f} (s, \cos\theta) = 8\pi \sum_{j = 0}^{\infty} (2j + 1) T_{i \rightarrow f}^{j}(s) \ d_{\lambda_f \lambda_i}^{j}(\theta) \ ,
\end{equation}
where $j$ is the total angular momentum of the final state (2-body) system,
$\lambda_i$ and $\lambda_f$ are the initial and final-state (2-body system)
helicities, $T_{i \rightarrow f}^{j}(s)$ are the amplitudes describing the
transition between the (definite helicity) states $i$ and $f$ for a given value
of $j$ and $d$ are the Wigner {\it {d}}--functions.
Multiplying both sides of the equation by
$d_{\lambda_f \lambda_i}^{j'}(\theta)$, integrating over $\cos\theta$ from $-1$
to $1$ and using the identity
\begin{equation}\label{eq:wigneridentity}
\int_{-1}^{1} {\rm d}\cos\theta \ d_{\lambda_f \lambda_i}^{j}(\theta) d_{\lambda_f \lambda_i}^{j'}(\theta) = \frac{2}{2j+1} \delta_{j' j} \ ,
\end{equation}
the $j$-th partial wave amplitude between the definite helicity states
$\lambda_i$ and $\lambda_f$ is given by
\begin{equation}\label{eq:jthpartialwave}
T_{i \rightarrow f}^{j}(s) = \frac{1}{16\pi} \int_{-1}^{1} {\rm d}\cos\theta {\cal{M}}_{i \rightarrow f} (s, \cos\theta) d_{\lambda_f \lambda_i}^{j}(\theta) \ .
\end{equation}
One therefore obtains,
\begin{equation}{\label{eqpartialwave}}
{\rm Im}(T^{j}_{i\to i} ) =  \sum_{f} \beta_{f}  |T^{j}_{i\to f}|^{2} = \beta_{i}|T^j_{i\to i}|^{2} +   \sum_{f \neq i}\beta_{f}|T^j_{i\to f}|^{2} \ ,
\end{equation}
which yields the following restrictions for the transition amplitudes
$T_{i \rightarrow f}^{j}(s)$,
\be\label{eq:unitarityconditions}
\beta_i {\rm Re}\left[ T_{i \rightarrow i}^{j}(s) \right] \leq 1\ , \qquad
\beta_i {\rm Im}\left[ T_{i \rightarrow i}^{j}(s) \right] \leq 2\ , \qquad
\sum_{f \neq i} \beta_i \beta_f \left| T_{i \rightarrow f}^{j}(s) \right|^2 \leq 1\ .
\ee

In order to compute the helicity amplitudes, we need explicit forms for the
wavefunctions of the external particles. We work in the Dirac representation throughout our calculation. Spinors of definite helicity $\lambda = \pm 1/2$,
propagating in the direction $(\theta,\phi)$ and describing particles with mass
$m$ and energy $E$ are represented as
\be
u(E, \theta,\phi, \lambda) = 
\left(
\begin{array}{c}
\sqrt{E + m} \ \chi_\lambda(\theta,\phi) \\
2\lambda \sqrt{E - m} \ \chi_\lambda(\theta,\phi)\\
\end{array}
\right)
\ \ \text{and}\ \
v(E, \theta,\phi, \lambda) = 
\left(
\begin{array}{c}
\sqrt{E - m} \ \chi_{-\lambda}(\theta,\phi) \\
-2\lambda \sqrt{E + m} \ \chi_{-\lambda}(\theta,\phi)\\
\end{array}
\right)\ ,
\ee
where the Weyl spinors $\chi$ are given by
\begin{equation}
\chi_{1/2} (\theta,\phi) = 
\left(
\begin{array}{c}
\cos\frac{\theta}{2} \\
e^{i\phi} \sin\frac{\theta}{2}\\
\end{array}
\right)
\qquad\text{and}\qquad
\chi_{-1/2} (\theta,\phi) = 
\left(
\begin{array}{c}
-e^{-i\phi} \sin\frac{\theta}{2} \\
\cos\frac{\theta}{2}\\
\end{array}
\right) \ .
\end{equation}
The conjugate spinors can be computed as usual with
$\bar{u} = u^\dagger \gamma^0$ and similarly for $\bar{v}$, with $\gamma^0$
being taken in the Dirac representation.
Polarisation vectors of massless vector fields are represented as
\begin{equation}
\epsilon_{\pm}^{\mu} = \frac{1}{\sqrt{2}} e^{\pm i\phi} 
\left( 0, \mp \cos\theta \cos\phi + i\sin\phi, \mp \cos\theta \sin\phi - i\cos\phi, \pm\sin\theta \right) \ ,
\end{equation}
and four-momenta are finally written as
\begin{equation}
p = \left( p, p\sin\theta \cos\phi, p\sin\theta \sin\phi, p\cos\theta \right)\ .
\end{equation}
The initial and final state helicities $\lambda_i$ and $\lambda_f$ appearing
in Eq.~\eqref{eq:jthpartialwave} are defined as
\mbox{$\lambda_i = \lambda_1 - \lambda_2$} and $\lambda_f = \lambda_3 - \lambda_4$, as we consider a $2 \to 2$
collision where the colliding particles are labelled as 1, 2, 3 and 4. By
convention, the particles 1 and 3 are chosen to propagate in the
$(\theta, \phi) = (0, 0)$ and $(\theta_f, 0)$ direction respectively, the choice
$\phi = 0$ not affecting the results since all distributions of final-state
particles are azimuthally symmetric. Consequently, the particles 2 and 4
propagate in the $(\pi - \theta, \pi + \phi) = (\pi, \pi)$ and
$(\pi - \theta_f, \pi)$ direction respectively.

For the new physics model considered in this paper, we treat the
momentum-dependent and momentum-independent operators separately.
Extracting the Feynman rules from the momentum-independent part of the
Lagrangian in Eq.~\eqref{eq:mdmonojets_lag}, the transition amplitude for the
$g g\to \eta\eta$ process reads
\begin{align}
{\cal{M}}_{\rm MI} =  \frac{\alpha_s c_{s\eta} c_{sg}}{4\pi} \epsilon_1^\mu \left(p_1^\mu p_2^\nu - g^{\mu\nu} (p_1 \cdot p_2)\right) \epsilon_2^\nu \frac{1}{k^2 - m_s^2} \ .
\end{align}
The only non-zero partial amplitudes are associated with the transition
\mbox{$(+,+) \rightarrow (0,0)$},
\begin{align}
T_{(+,+) \rightarrow (0,0)}^{0} = \frac{c_{sg} c_{s\eta} \alpha_s s}{64\pi^2 (m_s^2 - s)} \ ,
\end{align}
and similarly for $+ \leftrightarrow -$. Using the inequalities of
Eq.~\eqref{eq:unitarityconditions}, we get the following bound on
$\kappa_{\rm MI} = c_{s\eta} \times c_{sg}$,
\begin{equation}
\kappa_{MI} < \frac{64 \sqrt{2} \pi^2 (1 - \frac{m_s^2}{s})}{\alpha_s \left( 1 - \frac{4 m_\eta^2}{s}  \right)^{1/4}} \ ,
\end{equation}
where an extra factor of 1/2 has been added to account for the identical
final-state particles. For the momentum-dependent part of the Lagrangian, the transition amplitude
reads
\begin{align}
{\cal{M}}_{\rm MD} =  \frac{\alpha_s c_{s\eta} c_{sg}}{4\pi} \epsilon_1^\mu \left(p_1^\mu p_2^\nu - g^{\mu\nu} (p_1 \cdot p_2)\right) \epsilon_2^\nu \frac{k^2}{k^2 - m_s^2} \ .
\end{align}
Again, the only non-zero partial amplitudes are related to the
transition \mbox{$(+,+) \rightarrow (0,0)$},
\begin{align}
T_{(+,+) \rightarrow (0,0)}^{0} = \frac{c_{sg} c_{\partial s\eta} \alpha_s s^2}{64\pi^2 f^2 (m_s^2 - s)} \ ,
\end{align}
and similarly for $+ \leftrightarrow -$. We then extract a bound on
$\kappa_{\rm MD} = c_{\partial s\eta} \times c_{sg}$,
\begin{equation}
\kappa_{MD} < \frac{64 \sqrt{2} \pi^2 f^2 (s - m_s^2)}{\alpha_s s^2 \left( 1 - \frac{4 m_\eta^2}{s}  \right)^{1/4}} \ .
\end{equation}

Focusing on the process $gg \rightarrow \eta\eta$, and for given values of
masses and couplings, these relations can be used to extract the maximal allowed
value for $s$ for which our effective description makes sense perturbatively.
Conversely, for a given value of $s$ it can be used in order to bound the
parameters of our model, see Section~\ref{sec:compl}.

%% file: mdmonojets_paper.bbl
\providecommand{\href}[2]{#2}\begingroup\raggedright\begin{thebibliography}{10}

\bibitem{Aad:2014nra}
{\bf ATLAS} Collaboration, G.~Aad et~al., {\it {Search for pair-produced
  third-generation squarks decaying via charm quarks or in compressed
  supersymmetric scenarios in $pp$ collisions at $\sqrt{s}=8~$TeV with the
  ATLAS detector}},  {\em Phys. Rev.} {\bf D90} (2014), no.~5 052008,
  [\href{http://arxiv.org/abs/1407.0608}{{\tt arXiv:1407.0608}}].

\bibitem{Aad:2015zva}
{\bf ATLAS} Collaboration, G.~Aad et~al., {\it {Search for new phenomena in
  final states with an energetic jet and large missing transverse momentum in
  pp collisions at $\sqrt{s}=$8 TeV with the ATLAS detector}},  {\em Eur. Phys.
  J.} {\bf C75} (2015), no.~7 299, [\href{http://arxiv.org/abs/1502.01518}{{\tt
  arXiv:1502.01518}}]. [Erratum: Eur. Phys. J.C75,no.9,408(2015)].

\bibitem{Aaboud:2016tnv}
{\bf ATLAS} Collaboration, M.~Aaboud et~al., {\it {Search for new phenomena in
  final states with an energetic jet and large missing transverse momentum in
  $pp$ collisions at $\sqrt{s}=13$ TeV using the ATLAS detector}},
  \href{http://arxiv.org/abs/1604.07773}{{\tt arXiv:1604.07773}}.

\bibitem{Khachatryan:2014rra}
{\bf CMS} Collaboration, V.~Khachatryan et~al., {\it {Search for dark matter,
  extra dimensions, and unparticles in monojet events in proton-proton
  collisions at $\sqrt{s} = 8$ TeV}},  {\em Eur. Phys. J.} {\bf C75} (2015),
  no.~5 235, [\href{http://arxiv.org/abs/1408.3583}{{\tt arXiv:1408.3583}}].

\bibitem{Khachatryan:2015wza}
{\bf CMS} Collaboration, V.~Khachatryan et~al., {\it {Searches for
  third-generation squark production in fully hadronic final states in
  proton-proton collisions at $ \sqrt{s} = 8$ TeV}},  {\em JHEP} {\bf 06}
  (2015) 116, [\href{http://arxiv.org/abs/1503.08037}{{\tt arXiv:1503.08037}}].

\bibitem{Rizzo:2008fp}
T.~G. Rizzo, {\it {Identification of the Origin of Monojet Signatures at the
  LHC}},  {\em Phys. Lett.} {\bf B665} (2008) 361--368,
  [\href{http://arxiv.org/abs/0805.0281}{{\tt arXiv:0805.0281}}].

\bibitem{Buchmueller:2014yoa}
O.~Buchmueller, M.~J. Dolan, S.~A. Malik, and C.~McCabe, {\it {Characterising
  dark matter searches at colliders and direct detection experiments: Vector
  mediators}},  {\em JHEP} {\bf 01} (2015) 037,
  [\href{http://arxiv.org/abs/1407.8257}{{\tt arXiv:1407.8257}}].

\bibitem{Brooijmans:2016vro}
G.~Brooijmans et~al., {\it {Les Houches 2015: Physics at TeV colliders - new
  physics working group report}},  in {\em {9th Les Houches Workshop on Physics
  at TeV Colliders (PhysTeV 2015) Les Houches, France, June 1-19, 2015}}, 2016.
\newblock \href{http://arxiv.org/abs/1605.02684}{{\tt arXiv:1605.02684}}.

\bibitem{Frigerio:2012uc}
M.~Frigerio, A.~Pomarol, F.~Riva, and A.~Urbano, {\it {Composite Scalar Dark
  Matter}},  {\em JHEP} {\bf 07} (2012) 015,
  [\href{http://arxiv.org/abs/1204.2808}{{\tt arXiv:1204.2808}}].

\bibitem{Marzocca:2014msa}
D.~Marzocca and A.~Urbano, {\it {Composite Dark Matter and LHC Interplay}},
  {\em JHEP} {\bf 07} (2014) 107, [\href{http://arxiv.org/abs/1404.7419}{{\tt
  arXiv:1404.7419}}].

\bibitem{Fonseca:2015gva}
N.~Fonseca, R.~Z. Funchal, A.~Lessa, and L.~Lopez-Honorez, {\it {Dark Matter
  Constraints on Composite Higgs Models}},  {\em JHEP} {\bf 06} (2015) 154,
  [\href{http://arxiv.org/abs/1501.05957}{{\tt arXiv:1501.05957}}].

\bibitem{Brivio:2015kia}
I.~Brivio, M.~B. Gavela, L.~Merlo, K.~Mimasu, J.~M. No, R.~del Rey, and
  V.~Sanz, {\it {Non-linear Higgs portal to Dark Matter}},
  \href{http://arxiv.org/abs/1511.01099}{{\tt arXiv:1511.01099}}.

\bibitem{Bruggisser:2016nzw}
S.~Bruggisser, F.~Riva, and A.~Urbano, {\it {The Last Gasp of Dark Matter
  Effective Theory}},  \href{http://arxiv.org/abs/1607.02475}{{\tt
  arXiv:1607.02475}}.

\bibitem{Bruggisser:2016ixa}
S.~Bruggisser, F.~Riva, and A.~Urbano, {\it {Strongly Interacting Light Dark
  Matter}},  \href{http://arxiv.org/abs/1607.02474}{{\tt arXiv:1607.02474}}.

\bibitem{Aad:2015pla}
{\bf ATLAS} Collaboration, G.~Aad et~al., {\it {Constraints on new phenomena
  via Higgs boson couplings and invisible decays with the ATLAS detector}},
  {\em JHEP} {\bf 11} (2015) 206, [\href{http://arxiv.org/abs/1509.00672}{{\tt
  arXiv:1509.00672}}].

\bibitem{Khachatryan:2015vta}
{\bf CMS} Collaboration, V.~Khachatryan et~al., {\it {Search for exotic decays
  of a Higgs boson into undetectable particles and one or more photons}},  {\em
  Phys. Lett.} {\bf B753} (2016) 363--388,
  [\href{http://arxiv.org/abs/1507.00359}{{\tt arXiv:1507.00359}}].

\bibitem{Khachatryan:2014jba}
{\bf CMS} Collaboration, V.~Khachatryan et~al., {\it {Precise determination of
  the mass of the Higgs boson and tests of compatibility of its couplings with
  the standard model predictions using proton collisions at 7 and 8 $\,\text
  {TeV}$}},  {\em Eur. Phys. J.} {\bf C75} (2015), no.~5 212,
  [\href{http://arxiv.org/abs/1412.8662}{{\tt arXiv:1412.8662}}].

\bibitem{lacroix}
S.~Lacroix, {\it {Ph\'enom\'enologie dun candidat de mati\`ere noire coupl\'e
  au boson de Higgs}},  {\em
  https://phystev.cnrs.fr/wiki/\_media/2015:groups:higgs:dmhiggs:rapport.pdf}.

\bibitem{Craig:2014lda}
N.~Craig, H.~K. Lou, M.~McCullough, and A.~Thalapillil, {\it {The Higgs Portal
  Above Threshold}},  \href{http://arxiv.org/abs/1412.0258}{{\tt
  arXiv:1412.0258}}.

\bibitem{Panico:2015jxa}
G.~Panico and A.~Wulzer, {\it {The Composite Nambu-Goldstone Higgs}},  {\em
  Lect. Notes Phys.} {\bf 913} (2016) pp.1--316,
  [\href{http://arxiv.org/abs/1506.01961}{{\tt arXiv:1506.01961}}].

\bibitem{Alwall:2014bza}
J.~Alwall, C.~Duhr, B.~Fuks, O.~Mattelaer, D.~G. {\"O}zt{\"u}rk, and C.-H.
  Shen, {\it {Computing decay rates for new physics theories with FeynRules and
  MadGraph 5\_aMC@NLO}},  {\em Comput. Phys. Commun.} {\bf 197} (2015)
  312--323, [\href{http://arxiv.org/abs/1402.1178}{{\tt arXiv:1402.1178}}].

\bibitem{Alloul:2013bka}
A.~Alloul, N.~D. Christensen, C.~Degrande, C.~Duhr, and B.~Fuks, {\it
  {FeynRules 2.0 - A complete toolbox for tree-level phenomenology}},  {\em
  Comput. Phys. Commun.} {\bf 185} (2014) 2250--2300,
  [\href{http://arxiv.org/abs/1310.1921}{{\tt arXiv:1310.1921}}].

\bibitem{Alitti:1993pn}
{\bf UA2} Collaboration, J.~Alitti et~al., {\it {A Search for new intermediate
  vector mesons and excited quarks decaying to two jets at the CERN $\bar{p} p$
  collider}},  {\em Nucl. Phys.} {\bf B400} (1993) 3--24.

\bibitem{Aaltonen:2008dn}
{\bf CDF} Collaboration, T.~Aaltonen et~al., {\it {Search for new particles
  decaying into dijets in proton-antiproton collisions at s**(1/2) =
  1.96-TeV}},  {\em Phys. Rev.} {\bf D79} (2009) 112002,
  [\href{http://arxiv.org/abs/0812.4036}{{\tt arXiv:0812.4036}}].

\bibitem{Khachatryan:2015sja}
{\bf CMS} Collaboration, V.~Khachatryan et~al., {\it {Search for resonances and
  quantum black holes using dijet mass spectra in proton-proton collisions at
  $\sqrt{s} =$ 8 TeV}},  {\em Phys. Rev.} {\bf D91} (2015), no.~5 052009,
  [\href{http://arxiv.org/abs/1501.04198}{{\tt arXiv:1501.04198}}].

\bibitem{Aad:2014aqa}
{\bf ATLAS} Collaboration, G.~Aad et~al., {\it {Search for new phenomena in the
  dijet mass distribution using $p-p$ collision data at $\sqrt{s}=8$ TeV with
  the ATLAS detector}},  {\em Phys. Rev.} {\bf D91} (2015), no.~5 052007,
  [\href{http://arxiv.org/abs/1407.1376}{{\tt arXiv:1407.1376}}].

\bibitem{Belanger:2008sj}
G.~Belanger, F.~Boudjema, A.~Pukhov, and A.~Semenov, {\it {Dark matter direct
  detection rate in a generic model with micrOMEGAs 2.2}},  {\em Comput. Phys.
  Commun.} {\bf 180} (2009) 747--767,
  [\href{http://arxiv.org/abs/0803.2360}{{\tt arXiv:0803.2360}}].

\bibitem{Ade:2015xua}
{\bf Planck} Collaboration, P.~A.~R. Ade et~al., {\it {Planck 2015 results.
  XIII. Cosmological parameters}},  \href{http://arxiv.org/abs/1502.01589}{{\tt
  arXiv:1502.01589}}.

\bibitem{Feng:2005gj}
J.~L. Feng, S.~Su, and F.~Takayama, {\it {Lower limit on dark matter production
  at the large hadron collider}},  {\em Phys. Rev. Lett.} {\bf 96} (2006)
  151802, [\href{http://arxiv.org/abs/hep-ph/0503117}{{\tt hep-ph/0503117}}].

\bibitem{Busoni:2014gta}
G.~Busoni, A.~De~Simone, T.~Jacques, E.~Morgante, and A.~Riotto, {\it {Making
  the Most of the Relic Density for Dark Matter Searches at the LHC 14 TeV
  Run}},  {\em JCAP} {\bf 1503} (2015) 022,
  [\href{http://arxiv.org/abs/1410.7409}{{\tt arXiv:1410.7409}}].

\bibitem{Hisano:2010ct}
J.~Hisano, K.~Ishiwata, and N.~Nagata, {\it {Gluon contribution to the dark
  matter direct detection}},  {\em Phys. Rev.} {\bf D82} (2010) 115007,
  [\href{http://arxiv.org/abs/1007.2601}{{\tt arXiv:1007.2601}}].

\bibitem{Chu:2012qy}
X.~Chu, T.~Hambye, T.~Scarna, and M.~H.~G. Tytgat, {\it {What if Dark Matter
  Gamma-Ray Lines come with Gluon Lines?}},  {\em Phys. Rev.} {\bf D86} (2012)
  083521, [\href{http://arxiv.org/abs/1206.2279}{{\tt arXiv:1206.2279}}].

\bibitem{Shifman:1978zn}
M.~A. Shifman, A.~I. Vainshtein, and V.~I. Zakharov, {\it {Remarks on Higgs
  Boson Interactions with Nucleons}},  {\em Phys. Lett.} {\bf B78} (1978) 443.

\bibitem{Belanger:2014vza}
G.~Bélanger, F.~Boudjema, A.~Pukhov, and A.~Semenov, {\it {micrOMEGAs4.1: two
  dark matter candidates}},  {\em Comput. Phys. Commun.} {\bf 192} (2015)
  322--329, [\href{http://arxiv.org/abs/1407.6129}{{\tt arXiv:1407.6129}}].

\bibitem{Akerib:2015rjg}
{\bf LUX} Collaboration, D.~S. Akerib et~al., {\it {Improved WIMP scattering
  limits from the LUX experiment}},
  \href{http://arxiv.org/abs/1512.03506}{{\tt arXiv:1512.03506}}.

\bibitem{Akerib:2016vxi}
{\bf LUX} Collaboration, D.~S. Akerib et~al., {\it {Results from a search for
  dark matter in LUX with 332 live days of exposure}},
  \href{http://arxiv.org/abs/1608.07648}{{\tt arXiv:1608.07648}}.

\bibitem{Shoemaker:2011vi}
I.~M. Shoemaker and L.~Vecchi, {\it {Unitarity and Monojet Bounds on Models for
  DAMA, CoGeNT, and CRESST-II}},  {\em Phys. Rev.} {\bf D86} (2012) 015023,
  [\href{http://arxiv.org/abs/1112.5457}{{\tt arXiv:1112.5457}}].

\bibitem{Endo:2014mja}
M.~Endo and Y.~Yamamoto, {\it {Unitarity Bounds on Dark Matter Effective
  Interactions at LHC}},  {\em JHEP} {\bf 06} (2014) 126,
  [\href{http://arxiv.org/abs/1403.6610}{{\tt arXiv:1403.6610}}].

\bibitem{Kahlhoefer:2015bea}
F.~Kahlhoefer, K.~Schmidt-Hoberg, T.~Schwetz, and S.~Vogl, {\it {Implications
  of unitarity and gauge invariance for simplified dark matter models}},  {\em
  JHEP} {\bf 02} (2016) 016, [\href{http://arxiv.org/abs/1510.02110}{{\tt
  arXiv:1510.02110}}].

\bibitem{Conte:2012fm}
E.~Conte, B.~Fuks, and G.~Serret, {\it {MadAnalysis 5, A User-Friendly
  Framework for Collider Phenomenology}},  {\em Comput. Phys. Commun.} {\bf
  184} (2013) 222--256, [\href{http://arxiv.org/abs/1206.1599}{{\tt
  arXiv:1206.1599}}].

\bibitem{Conte:2014zja}
E.~Conte, B.~Dumont, B.~Fuks, and C.~Wymant, {\it {Designing and recasting LHC
  analyses with MadAnalysis 5}},  {\em Eur. Phys. J.} {\bf C74} (2014), no.~10
  3103, [\href{http://arxiv.org/abs/1405.3982}{{\tt arXiv:1405.3982}}].

\bibitem{13tevmonojet}
D.~Sengupta, {\it {Madanalysis5 implementation of the ATLAS monojet and missing
  transverse momentum search documented in arXiv:1604.07773}},
  10.7484/INSPIREHEP.DATA.GTH3.RN26.

\bibitem{Dumont:2014tja}
B.~Dumont, B.~Fuks, S.~Kraml, S.~Bein, G.~Chalons, E.~Conte, S.~Kulkarni,
  D.~Sengupta, and C.~Wymant, {\it {Toward a public analysis database for LHC
  new physics searches using MADANALYSIS 5}},  {\em Eur. Phys. J.} {\bf C75}
  (2015), no.~2 56, [\href{http://arxiv.org/abs/1407.3278}{{\tt
  arXiv:1407.3278}}].

\bibitem{Degrande:2011ua}
C.~Degrande, C.~Duhr, B.~Fuks, D.~Grellscheid, O.~Mattelaer, and T.~Reiter,
  {\it {UFO - The Universal FeynRules Output}},  {\em Comput. Phys. Commun.}
  {\bf 183} (2012) 1201--1214, [\href{http://arxiv.org/abs/1108.2040}{{\tt
  arXiv:1108.2040}}].

\bibitem{Alwall:2014hca}
J.~Alwall, R.~Frederix, S.~Frixione, V.~Hirschi, F.~Maltoni, O.~Mattelaer,
  H.~S. Shao, T.~Stelzer, P.~Torrielli, and M.~Zaro, {\it {The automated
  computation of tree-level and next-to-leading order differential cross
  sections, and their matching to parton shower simulations}},  {\em JHEP} {\bf
  07} (2014) 079, [\href{http://arxiv.org/abs/1405.0301}{{\tt
  arXiv:1405.0301}}].

\bibitem{Sjostrand:2006za}
T.~Sj{\"o}strand, S.~Mrenna, and P.~Z. Skands, {\it {PYTHIA 6.4 Physics and
  Manual}},  {\em JHEP} {\bf 05} (2006) 026,
  [\href{http://arxiv.org/abs/hep-ph/0603175}{{\tt hep-ph/0603175}}].

\bibitem{deFavereau:2013fsa}
{\bf DELPHES 3} Collaboration, J.~de~Favereau, C.~Delaere, P.~Demin,
  A.~Giammanco, V.~Lemaitre, A.~Mertens, and M.~Selvaggi, {\it {DELPHES 3, A
  modular framework for fast simulation of a generic collider experiment}},
  {\em JHEP} {\bf 02} (2014) 057, [\href{http://arxiv.org/abs/1307.6346}{{\tt
  arXiv:1307.6346}}].

\bibitem{Cacciari:2011ma}
M.~Cacciari, G.~P. Salam, and G.~Soyez, {\it {FastJet User Manual}},  {\em Eur.
  Phys. J.} {\bf C72} (2012) 1896, [\href{http://arxiv.org/abs/1111.6097}{{\tt
  arXiv:1111.6097}}].

\bibitem{Cacciari:2008gp}
M.~Cacciari, G.~P. Salam, and G.~Soyez, {\it {The Anti-k(t) jet clustering
  algorithm}},  {\em JHEP} {\bf 04} (2008) 063,
  [\href{http://arxiv.org/abs/0802.1189}{{\tt arXiv:0802.1189}}].

\bibitem{Read:2000ru}
A.~L. Read, {\it {Modified frequentist analysis of search results (The CL(s)
  method)}},  in {\em {Workshop on confidence limits, CERN, Geneva,
  Switzerland, 17-18 Jan 2000: Proceedings}}, 2000.

\bibitem{Read:2002hq}
A.~L. Read, {\it {Presentation of search results: The CL(s) technique}},  {\em
  J. Phys.} {\bf G28} (2002) 2693--2704. [,11(2002)].

\bibitem{Sjostrand:2014zea}
T.~Sjstrand, S.~Ask, J.~R. Christiansen, R.~Corke, N.~Desai, P.~Ilten,
  S.~Mrenna, S.~Prestel, C.~O. Rasmussen, and P.~Z. Skands, {\it {An
  Introduction to PYTHIA 8.2}},  {\em Comput. Phys. Commun.} {\bf 191} (2015)
  159--177, [\href{http://arxiv.org/abs/1410.3012}{{\tt arXiv:1410.3012}}].

\bibitem{Aad:2009wy}
{\bf ATLAS} Collaboration, {\it {Expected Performance of the ATLAS Experiment -
  Detector, Trigger and Physics}},  {\em SLAC-R-980, CERN-OPEN-2008-020} (2009)
  [\href{http://arxiv.org/abs/0901.0512}{{\tt arXiv:0901.0512}}].

\bibitem{Moneta:2010pm}
L.~Moneta, K.~Belasco, K.~S. Cranmer, S.~Kreiss, A.~Lazzaro, D.~Piparo,
  G.~Schott, W.~Verkerke, and M.~Wolf, {\it {The RooStats Project}},  {\em PoS}
  {\bf ACAT2010} (2010) 057, [\href{http://arxiv.org/abs/1009.1003}{{\tt
  arXiv:1009.1003}}].

\bibitem{Abercrombie:2015wmb}
D.~Abercrombie et~al., {\it {Dark Matter Benchmark Models for Early LHC Run-2
  Searches: Report of the ATLAS/CMS Dark Matter Forum}},
  \href{http://arxiv.org/abs/1507.00966}{{\tt arXiv:1507.00966}}.

\bibitem{Arina:2016cqj}
C.~Arina et~al., {\it {A comprehensive approach to dark matter studies:
  exploration of simplified top-philic models}},
  \href{http://arxiv.org/abs/1605.09242}{{\tt arXiv:1605.09242}}.

\bibitem{Jacob:1959at}
M.~Jacob and G.~C. Wick, {\it {On the general theory of collisions for
  particles with spin}},  {\em Annals Phys.} {\bf 7} (1959) 404--428. [Annals
  Phys.281,774(2000)].

\bibitem{Haber:1994pe}
H.~E. Haber, {\it {Spin formalism and applications to new physics searches}},
  in {\em {Spin structure in high-energy processes: Proceedings, 21st SLAC
  Summer Institute on Particle Physics, 26 Jul - 6 Aug 1993, Stanford, CA}},
  1994.
\newblock \href{http://arxiv.org/abs/hep-ph/9405376}{{\tt hep-ph/9405376}}.

\end{thebibliography}\endgroup
